INSTITUTO TECNOLÓGICO AUTÓNOMO DE MÉXICO

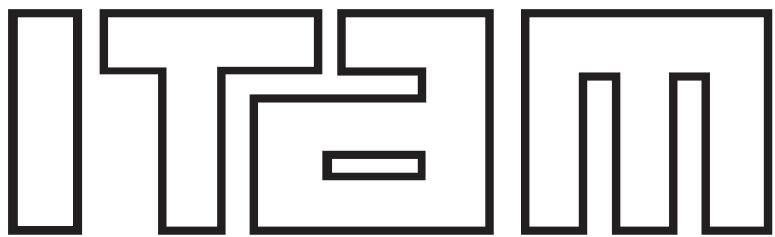

ANÁLISIS DE COINTEGRACIÓN CON UNA APLICACIÓN AL MERCADO DE DEUDA EN ESTADOS UNIDOS, CANADÁ Y MÉXICO

# TESINA

QUE PARA OBTENER EL TÍTULO DE:
LICENCIADO EN MATEMATICAS APLICADAS

P R E S E N T A :

**EMILIANO DÍAZ SALAS-PORRAS**

**ASESOR:**
DR. ALEJANDRO ISLAS CAMARGO

**REVISOR:**
DR. VÍCTOR MANUEL GUERRERO GUZMÁN

**MÉXICO, D.F.**                                   **2014**

**CONTENIDO**

# 1 Introducción

## 1.1 Objetivo

Revisar y resumir ciertos aspectos relevantes de la teoría de Vectores Autorregresivos (VAR) como herramientas para modelar series de tiempo económicas, en específico su capacidad para incluir información tanto de corto plazo como de largo plazo. Se derivará el modelo VAR en su forma Modelo de Corrección de Errores y se estudiará la descomposición en componentes permanentes y transitorias propuesta por Gonzalo y Granger (1995). Se dará una exposición introductoria a la teoría de estimación de modelos de regresión de rango reducido, necesaria para estimar el modelo VAR en su forma MCE.

Se realizará una aplicación de análisis de cointegración en la que se analizarán las tasas de interés de Estados Unidos, México y Canadá a distintos plazos, con el objetivo de encontrar los factores comunes de largo plazo que impulsan al sistema. Se usarán las técnicas VAR mencionadas, en particular se estimará, mediante el procedimiento de Johansen, el VAR correspondiente al sistema, en su forma MCE. A partir de esta estimación se calculará la descomposición del VAR en componentes permanentes y transitorios. Se realizarán pruebas de hipótesis sobre los componentes permanentes para identificar qué tasas, de las nueve estudiadas, son las que impulsan el sistema.

Se han considerado tres tasas de interés en cada país, cada cual con sus propios plazos de vencimiento: corto, mediano y largo plazo. En el caso de Estados Unidos las tasas de corto, mediano y largo plazo corresponden al rendimiento de mercado de títulos de la tesorería con plazos a vencimiento constante de 3 meses, 3 años y 10 años respectivamente. En el caso de Canadá la tasa de corto plazo corresponde al rendimiento promedio de títulos de la tesorería con plazo de 3 meses. Las tasas de mediano y largo plazo se refieren a la tasa de bonos gubernamentales con plazo a vencimiento original de 3-5 años y de 10 años o más respectivamente. Para México la tasa de corto plazo corresponde al rendimiento promedio de títulos de la tesorería con plazo de 3 meses (CETES 91 días). La tasa de mediano y largo plazo se refiere a la tasa promedio de bonos gubernamentales con plazo a vencimiento original de 3 y 10 años respectivamente.

En todos los casos se analizaron tasas nominales anualizadas. La periodicidad de las series de tiempo es mensual y el periodo analizado corre de enero de 2002 hasta diciembre de 2013, es decir, cada serie cuenta con 144 observaciones mensuales.

La aplicación de análisis de cointegración a los mercados de deuda de Norteamérica está basado en el estudio de Gonzalo y Granger (1995). En aquel estudio, se aplicaron las técnicas mencionadas para analizar las tasas de corto, mediano y largo plazo de Estados Unidos y Canadá. En aquel entonces, se analizaron las mismas tasas de interés de Estados Unidos y Canadá pero para el periodo que corre de enero de 1969 a diciembre de 1988. En dicho estudio se concluyó que existía un solo factor permanente que explicaba el comportamiento de largo plazo de todo el sistema. Adicionalmente, a partir de una prueba de hipótesis sobre el factor permanente, se probó que ese factor se componía únicamente de las tasas de interés de Estados Unidos. Esto llevó a la conclusión que el comportamiento de largo plazo del sistema era determinado por lo que sucedía en Estados Unidos. En este trabajo interesa corroborar si el sistema sigue teniendo el mismo grado de cointegración y si las tasas de Estados Unidos siguen siendo el único factor que impulsa al sistema. Se ha ampliado el "sistema" analizado para incluir a México.

### 1.2  *Justificación para el uso del Modelo VAR*

El modelo VAR resulta ser una herramienta atractiva para modelar series macroeconómicas dado que:

- **Es un modelo flexible**: permite incorporar información de distinto tipo; estocástica o determinista; en escala ordinal, nominal o continua; y en general, permite describir diversos sistemas económicos.
- **Sus parámetros son fáciles de estimar**: los estimadores máximo verosímiles se pueden calcular analíticamente.
- **Adecuación a series económicas**: en muchos casos, estos modelos ajustan bien a series de tiempo económicas al modelar de forma adecuada la fuerte dependencia temporal que las caracteriza.
- **Incorpora información de corto y largo plazo:** al incluir información sobre relaciones de corto y largo plazo, es posible modelar relaciones de equilibrio de largo plazo, tendencias comunes, interacciones entre las series y retroalimentación entre distintos procesos.

Pocas variables macroeconómicas se pueden considerar fijas o no estocásticas. Siguiendo el análisis de Johansen y Juselius (2006), un acercamiento econométrico que reconociera esto implica una formulación probabilística de todo el proceso generador de datos. En contraste, los modelos basados en teoría económica, tradicionalmente se han desarrollado suponiendo que las variables explicativas son fijas o predeterminadas. Al aplicarse a datos empíricos, se agrega un error estocástico al modelo matemático para explicar la falta de ajuste del modelo. A continuación se mostrará que, bajo ciertos supuestos, el modelo VAR puede representar una formulación probabilística de todo el proceso generador de las distintas variables analizadas en un estudio econométrico.

De la forma más general, podemos representar las variables macroeconómicas de interés como un vector de variables aleatorias que son parte de un proceso estocástico:

$$Z = \begin{bmatrix} X_1 \\ X_2 \\ \vdots \\ X_T \end{bmatrix} \qquad X_t = \begin{bmatrix} x_{1t} \\ x_{2t} \\ \vdots \\ x_{pt} \end{bmatrix}$$

Donde $p$ representa el número de componentes o variables de interés, $T$ representa el número de realizaciones del proceso estocástico y $t \in \{1, 2, \ldots, T\}$ indica una realización en el momento $t$ del proceso estocástico.

Suponemos que cada $X_t$ es una normal multivariada, $X_t \sim N(\mu_t, \Sigma_t)$, de tal forma que estimando los primeros dos momentos correctamente, conocemos la función de densidad conjunta.

$$E[Z] = \mu = \begin{bmatrix} \mu_1 \\ \mu_2 \\ \vdots \\ \mu_T \end{bmatrix} \epsilon \, \mathbb{R}^{Tp} \qquad \Sigma = E[(Z-\mu)(Z-\mu)'] = \begin{bmatrix} \Sigma_{11} & \Sigma_{21} & \cdots & \Sigma_{T1} \\ \Sigma_{21} & \Sigma_{22} & \cdots & \Sigma_{T2} \\ \vdots & \vdots & \ddots & \vdots \\ \Sigma_{T1} & \Sigma_{T2} & \cdots & \Sigma_{TT} \end{bmatrix} \epsilon \, \mathbb{R}^{TpxTp}$$

en donde $\Sigma_{t\tau} = Cov(X_t, X_\tau) = Cov(X_\tau, X_t) = \Sigma_{\tau t}$. Re-indexamos la matriz de covarianzas en términos de los retrasos, de tal forma que $\Sigma_{th} = Cov(X_t, X_{t-h})$ y

$$\Sigma = \begin{bmatrix} \Sigma_{10} & \Sigma_{21} & \cdots & \Sigma_{T,T-1} \\ \Sigma_{21} & \Sigma_{20} & \cdots & \Sigma_{T,T-2} \\ \vdots & \vdots & \ddots & \vdots \\ \Sigma_{T,T-1} & \Sigma_{T,T-2} & \cdots & \Sigma_{T0} \end{bmatrix}$$

La situación habitual es que para cada vector aleatorio $X_t$ únicamente se cuenta con una observación por lo que en total se cuenta con $T$ observaciones. En contraste se puede verificar que, con el supuesto de normalidad, tenemos un modelo con $Tp + Tp^2 \left[\frac{T+1}{2}\right]$ parámetros. Para reducir el número de nuestro modelo de tal forma que sea posible realizar inferencia estadística sobre él, realizamos los dos siguientes supuestos:

1. $\mu_t = \mu \quad \forall\, t \in \{1, \dots, T\}$
2. $\Sigma_{th} = \Sigma_h \,\forall\, t \in \{1, \dots, T\}, h \in \{\dots, -1, 0, 1, \dots\}$

Estos supuestos garantizan que los parámetros del modelo son los mismos, independientemente del periodo en el que se observa el proceso estocástico. Así mismo, definen un proceso débilmente estacionario. Con esta simplificación, y dividiendo el vector en dos sub-vectores, tenemos lo siguiente:

$$Z = \begin{bmatrix} X_T \\ X_{T-1}^0 \end{bmatrix}, \quad X_{T-1}^0 = \begin{bmatrix} X_{T-1} \\ X_{T-2} \\ \vdots \\ X_1 \end{bmatrix}, \quad E[X_T] = m_1,\ E[X_{T-1}^0] = m_2,$$

$$\Sigma = \begin{bmatrix} \Sigma_0 & \Sigma_1 & \cdots & \Sigma_{T-1} \\ \Sigma_1 & \Sigma_0 & \cdots & \Sigma_{T-2} \\ \vdots & \vdots & \ddots & \vdots \\ \Sigma_{T-1} & \Sigma_{T-2} & \cdots & \Sigma_0 \end{bmatrix} = \begin{bmatrix} \Lambda_{11} & \Lambda_{12} \\ \Lambda_{21} & \Lambda_{22} \end{bmatrix}, \Lambda_{11} = \Sigma_0, \Lambda_{21} = \begin{bmatrix} \Sigma_1 \\ \Sigma_2 \\ \vdots \\ \Sigma_{T-1} \end{bmatrix} = \Lambda_{12}', \quad y$$

$$\Lambda_{22} = \begin{bmatrix} \Sigma_0 & \Sigma_1 & \cdots & \Sigma_{T-2} \\ \Sigma_1 & \Sigma_0 & \cdots & \Sigma_{T-3} \\ \vdots & \vdots & \vdots & \vdots \\ \Sigma_{T-2} & \Sigma_{T-3} & \cdots & \Sigma_0 \end{bmatrix}$$

Como $Z$ es un vector aleatorio normal multivariado sabemos que $X_T | X_{T-1}^0$ también es una normal multivariada con vector de esperanza $\mathrm{E}[X_T | X_{T-1}^0] = m_{1.2}$ y matriz de varianzas y covarianzas $\mathrm{Var}[X_T | X_{T-1}^0] = \Lambda_{1.2}$. Derivemos la forma de $m_{1.2}$. Considerar el vector $z = X_T + A\, X_{T-1}^0$ con $A = -\Lambda_{12} \Lambda_{22}^{-1}$ y ver que $\mathrm{Cov}[z, X_{T-1}^0] = 0$:

$\text{Cov}[z, X_{T-1}^0] = \text{Cov}[X_T + A\, X_{T-1}^0, X_{T-1}^0] = \text{Cov}[X_T, X_{T-1}^0] + A\text{Var}[X_{T-1}^0] = \Lambda_{12} - \Lambda_{12}\Lambda_{22}^{-1}\Lambda_{22} = 0$

Como $z$ y $X_{T-1}^0$ son normales y no están correlacionadas, son independientes. Por lo tanto:

$$\begin{aligned}m_{1.2} = E[X_T|X_{T-1}^0] &= E[z - A\, X_{T-1}^0|X_{T-1}^0] = E[z|X_{T-1}^0] - AE[X_{T-1}^0|X_{T-1}^0] = E[z] - AX_{T-1}^0 \\ &= E[X_T + A\, X_{T-1}^0] - AX_{T-1}^0 = E[X_T] + A\, E[X_{T-1}^0] - AX_{T-1}^0 \\ &= m_1 + A(m_2 - X_{T-1}^0) = m_1 + \Lambda_{12}\Lambda_{22}^{-1}(X_{T-1}^0 - m_2)\end{aligned}$$

De ahí que si definimos $\varepsilon_T = X_T - m_{1.2}$ entonces:

$$\begin{aligned}X_T = m_{1.2} + \varepsilon_T &= m_1 + \Lambda_{12}\Lambda_{22}^{-1}(X_{T-1}^0 - m_2) + \varepsilon_T \\ &= (m_1 - \Lambda_{12}\Lambda_{22}^{-1}m_2) + (\Lambda_{12}\Lambda_{22}^{-1})X_{T-1}^0 + \varepsilon_T\end{aligned}$$

Donde,
- $m_1 - \Lambda_{12}\Lambda_{22}^{-1}m_2 \in \mathbb{R}^p$
- $\Lambda_{12}\Lambda_{22}^{-1} \in \mathbb{R}^{p \times (p(T-1))}$

Redefiniendo $\mu_0 = m_1 - \Lambda_{12}\Lambda_{22}^{-1}m_2$, $[\Pi_1, \Pi_2, \ldots, \Pi_{T-1}] = \Lambda_{12}\Lambda_{22}^{-1}$ con $\Pi_i \in \mathbb{R}^{p \times p}$ y $\Sigma = \Lambda_{1.2} \in \mathbb{R}^{p \times p}$ obtenemos la representación VAR de los datos:

$X_T = \mu_0 + \Pi_1 X_{T-1} + \Pi_2 X_{T-2} + \cdots + \Pi_k X_{T-k} + \varepsilon_T, \ \varepsilon_T \sim N_p(0, \Sigma)$, iid

En conclusión, si se cumple que el vector aleatorio sigue una distribución normal y que la esperanza y covarianza de sus componentes no varía con respecto al tiempo (es un proceso estacionario) entonces, el VAR resulta ser una formulación probabilística adecuada para modelar el proceso que genera los datos. En este caso, y como se muestra en el desarrollo anterior, el VAR resulta ser una reformulación de la covarianza de los datos.

En la siguiente sección se mostrará que el modelo VAR permite incluir información de corto y largo plazo. En contraste, al aplicar ciertas técnicas univariadas, se pierde la información de largo plazo. Esto debido a que estas técnicas requieren de series estacionarias, pero al quitarle sus tendencias, determinísticas o estocásticas, a los datos se pierde la información de largo plazo.

## 2   Modelo de Corrección de Errores

### *2.1   Forma reducida y representación MCE*

Suponer que $X_t$ es un vector con $p$ series de tiempo autorregresivas de orden $k$ que están integradas de orden 1:

$$(X_t)_i \sim I(1) \; \forall \; i \; \epsilon \; \{1,2, \dots, p\}$$

A continuación se muestra la representación VAR, del vector $X_t$:

$$X_t = \Pi_0 + \Pi_1 X_{t-1} + \Pi_2 X_{t-2} + \cdots + \Pi_k X_{t-k} + \varepsilon_t$$

Donde,
- $X_t, \; \Pi_0, \; \varepsilon_t \in \mathbb{R}^p$
- $\Pi_i \in \mathbb{R}^{pxp}$
- $\varepsilon_t$ es una serie estacionaria

Sumando y restando $\Pi_i X_{t-1}$, para toda $i \in \{2,3, \dots, k\}$, se obtiene la siguiente expresión:

$$X_t = \Pi_0 + (\Pi_1 + \Pi_2 + \cdots + \Pi_k)X_{t-1} + \Pi_2(X_{t-2} - X_{t-1}) + \cdots + \Pi_k(X_{t-k} - X_{t-1}) + \varepsilon_t$$

Como
$$X_{t-1} - X_{t-j} = X_{t-1} - (X_{t-2} - X_{t-2}) - (X_{t-3} - X_{t-3}) - \cdots - (X_{t-j+1} - X_{t-j+1}) - X_{t-j}$$
$$= \nabla X_{t-1} + \nabla X_{t-2} + \cdots + \nabla X_{t-j+1} = \sum_{i=1}^{j-1} \nabla X_{t-i}$$

Tenemos que:

$$X_t = \Pi_0 + (\Pi_1 + \Pi_2 + \cdots + \Pi_k)X_{t-1} - \Pi_2 \nabla X_{t-1} - \Pi_3(\nabla X_{t-1} + \nabla X_{t-2}) - \cdots$$
$$- \Pi_k(\nabla X_{t-1} + \nabla X_{t-2} + \cdots \nabla X_{t-k+1}) + \varepsilon_t$$

Reagrupando términos tenemos que:

$$X_t = \Pi_0 + (\Pi_1 + \Pi_2 + \cdots + \Pi_k)X_{t-1} - (\Pi_2 + \Pi_3 + \cdots + \Pi_k)\nabla X_{t-1} - (\Pi_3 + \cdots + \Pi_k)\nabla X_{t-2}$$
$$- \cdots - (\Pi_3 + \cdots + \Pi_k)\nabla X_{t-k+1} + \varepsilon_t$$

Restando $X_{t-1}$ de ambos lados y usando las siguientes identidades,

$$\Gamma_0 = \Pi_0$$

$$\Gamma = -(I - (\Pi_1 + \Pi_2 + \cdots + \Pi_k)) = -\left(I - \sum_{i=1}^{k} \Pi_i\right)$$

$$\Gamma_i = -(\Pi_{i+1} + \Pi_{i+2} + \cdots + \Pi_k) = -\sum_{j=i+1}^{k} \Pi_j$$

Obtenemos la representación de $X_t$ como Modelo de Corrección de Errores (MCE):

$$\nabla X_t = \Gamma_0 + \Gamma X_{t-1} - \Gamma_1 \nabla X_{t-1} - \Gamma_2 \nabla X_{t-2} - \cdots - \Gamma_k \nabla X_{t-k+1} + \varepsilon_t$$

Como $\nabla X_t$ y $\varepsilon_t$ son series estacionarias esto significa que $\Gamma X_t$ debe de ser estacionaria para mantener la consistencia de la representación MCE. Si $\text{rango}(\Gamma) = p$, esto significa que cualquier combinación lineal de $X_t$ es estacionaria lo que es contradictorio por lo que:

$$\text{rango}(\Gamma) = r$$
$$r \in \{0, 1, \ldots, p-1\}$$

**Definición 1 (Espacio de Cointegración)**: Si $\Gamma_{*i}$ es la columna $i$ de $\Gamma$, el **Espacio de Cointegración** se define como:

$$\mathcal{C} \coloneqq span\{\Gamma_{*1}, \Gamma_{*2}, \ldots, \Gamma_{*p}\} \subset \mathbb{R}^p$$

Si, por ejemplo, tomamos como base de $\mathcal{C}$ las primeras $r$ columnas linealmente independientes de $\Gamma$ para construir una matriz $\alpha \in \mathbb{R}^{p \times r}$ queda claro que existe una factorización $\Gamma = \alpha \beta'$ tal que:

1. $\alpha, \beta \in \mathbb{R}^{p \times r}$
2. $\alpha, \beta$ de rango completo por columna
3. $\beta' X_t \sim I(0)$

Esta factorización no es única. Considerar una matriz Q invertible. En este caso la factorización $\Gamma = ab' = (\alpha Q)(\beta(Q')^{-1})' = \alpha Q Q^{-1}\beta' = \alpha\beta'$ también cumple las condiciones 1 y 2. Será necesario imponer ciertas condiciones de normalización para poder identificar a $\alpha$ y $\beta$ de forma única. La dimensión del espacio de cointegración es $r$. Si la dimensión del espacio de cointegración es $r > 0$ se dice que el vector $X_t$ está cointegrado.

**Definición 2. Espacio de Tendencias Comunes:** El **Espacio de Tendencias Comunes** se define como:

$$\mathcal{T} \coloneqq \mathbb{R}^p \setminus \mathcal{C}$$

La dimensión del espacio de tendencias comunes es $k = p - r \in \{1,2,\ldots,p\}$.

## 2.2 Descomposición P-T Gonzalo-Granger

Siguiendo el tratamiento de Gonzalo y Granger (1995) si el espacio de cointegración no es nulo ($r > 0$) entonces el vector $X_t$ es integrado de orden 1 y se puede expresar como la suma de un componente vectorial integrado de orden 1 y un componente vectorial integrado de orden 0. Como la dimensión del espacio de tendencias comunes es $k$, $X_t$ se puede explicar a partir de combinaciones lineales de $k < p$ factores. Notar que aquí, $f_t$ no necesariamente es función lineal de $X_t$.

$$X_t = A_1 f_t + \widetilde{X_t}$$

Donde
1. $A_1 \epsilon \mathbb{R}^{pxk}$
2. $f_t, \widetilde{X_t} \epsilon \mathbb{R}^{px1}$
3. $f_t \sim I(1)$
4. $\widetilde{X_t}$ es estacionaria

**Definición 3. (Descomposición Permanente-Transitorio de Gonzalo-Granger).** Sea $X_t$ una serie de tiempo estacionaria en sus diferencias, es decir integrada de orden 1. Entonces una descomposición Permanente-Transitorio (P-T) para $X_t$ es un par de series de tiempo, $P_t$ y $T_t$, tal que:

1. $P_t$ es estacionaria en sus diferencias y $T_t$ es estacionaria
2. $Var(\nabla P_t)$ y $Var(T_t) > 0$
3. $X_t = P_t + T_t$
4. Si $H(L)\begin{bmatrix}\nabla P_t \\ T_t\end{bmatrix} = \begin{bmatrix}u_{Pt} \\ u_{Tt}\end{bmatrix}$ es la representación autorregresiva del vector $(\nabla P_t, T_t)$, con $u_{Pt}$ y $u_{Tt}$, no correlacionados (VAR en su forma estructural) entonces
    a. $\lim_{h\to\infty}\dfrac{\partial E_t[X_{t+h}]}{u_{Pt}} \neq 0$
    b. $\lim_{h\to\infty}\dfrac{\partial E_t[X_{t+h}]}{u_{Tt}} = 0$

Esta última condición indica que para que la descomposición $X_t = P_t + T_t$ sea P-T en el sentido de Gonzalo-Granger, la esperanza del proceso $X_t$ debe ser sensible, a la larga, ante choques en el componente permanente $P_t$ e insensible, a la larga, ante choques en el componente transitorio $T_t$.

Para identificar el componente permanente $P_t = A_1 f_t$ y el componente transitorio $T_t = \widetilde{X_t}$, Gonzalo y Granger proponen realizar dos supuestos:

1. $f_t = BX_t$, con $B \in \mathbb{R}^{k \times p}$, es decir que los factores permanentes sean función lineal de las series de tiempo observadas, y
2. Que la descomposición $X_t = A_1 f_t + \widetilde{X_t}$ es una descomposición P-T en el sentido de la definición anterior.

Sustituyendo la ecuación del primer supuesto en la descomposición tenemos:

$$X_t = A_1 B X_t + \widetilde{X_t} \Rightarrow \widetilde{X_t} = (I_p - A_1 B)X_t = CX_t$$

Como $rango(C) \leq r$ y $rango(\beta) = r$ podemos descomponer la matriz $C$ de la siguiente forma:

$$C = I - A_1 B = A_2 \beta'$$

donde $A_2 \in \mathbb{R}^{pxr}$. Es decir que podemos reexpresar la descomposición de la siguiente forma:

$$X_t = A_1 f_t + A_2 \beta' X_t = A_1 f_t + A_2 z_t$$

Para deducir la forma de $f_t$ regresamos al VAR en su forma MCE :

$$\nabla X_t = \Gamma_0 + \alpha\beta' X_{t-1} - \Gamma_1 \nabla X_{t-1} - \Gamma_2 \nabla X_{t-2} - \cdots - \Gamma_k \nabla X_{t-k+1} + \varepsilon_t$$

Buscamos encontrar combinaciones lineales $f_t = BX_t$ tal que el componente transitorio $\widetilde{X_t} = A_2\beta' X_t$ no tenga impacto de largo plazo sobre $X_t$. Si multiplicamos el MCE por $\alpha_\perp \in \mathbb{R}^{pxk}$, tal que $\alpha'_\perp \alpha = 0$, se obtiene:

$$\alpha'_\perp \nabla X_t = \alpha'_\perp \Gamma_0 + \alpha'_\perp \alpha\beta' X_{t-1} - \alpha'_\perp \Gamma_i \nabla X_{t-1} - \cdots - \alpha'_\perp \Gamma_k \nabla X_{t-k+1} + \alpha'_\perp \varepsilon_t$$
$$= \alpha'_\perp \Gamma_0 - \alpha'_\perp \Gamma_i \nabla X_{t-1} - \cdots - \alpha'_\perp \Gamma_k \nabla X_{t-k+1} + \alpha'_\perp \varepsilon_t$$

De tal forma que $\widetilde{X_t}$ no tiene impacto de largo plazo sobre $X_t$. En otras palabras tenemos que:
$$f_t = \alpha'_\perp X_t$$

Reformulando una vez más la descomposición P-T que buscamos, se tiene que:

$$X_t = A_1 \alpha'_\perp X_t + A_2 \beta' X_t$$

Ahora considerar $x \in \mathbb{R}^p$. Podemos expresar a $x$ de las siguientes formas:

$$x = \alpha a + \alpha_\perp b = \beta c + \beta_\perp d$$

Donde $a, c \in \mathbb{R}^r$ y $b, d \in \mathbb{R}^k$. Pero por la descomposición P-T se requiere que :

$$I_p = A_1 \alpha'_\perp + A_2 \beta'$$

De ahí que:

$$x = \alpha a + \alpha_\perp b = (A_1 \alpha'_\perp + A_2 \beta')(\alpha a + \alpha_\perp b) = (A_2 \beta' \alpha)a + (A_1 \alpha'_\perp \alpha + A_2 \beta' \alpha_\perp)b$$
$$\Rightarrow A_2 \beta' = I \Rightarrow A_2(\beta'\alpha) = \alpha$$

y

$$x = \beta c + \beta_\perp d = (A_1 \alpha'_\perp + A_2 \beta')(\beta c + \beta_\perp d) = (A_1 \alpha'_\perp \beta + A_2 \beta' \beta)c + (A_1 \alpha'_\perp \beta_\perp)d$$
$$\Rightarrow A_1 \alpha'_\perp = I \Rightarrow A_1(\alpha'_\perp \beta_\perp) = \beta_\perp$$

Como $\beta'\alpha \in \mathbb{R}^{rxr}$, $\alpha'_\perp \beta_\perp \in \mathbb{R}^{kxk}$, $rango(\beta'\alpha) = r$ y $rango(\alpha'_\perp \beta_\perp) = k$ tenemos que:

$$A_1 = \beta_\perp (\alpha'_\perp \beta_\perp)^{-1}$$
$$A_2 = \alpha(\beta'\alpha)^{-1}$$

## *2.3 Estimación*

En esta sección, a modo de introducción a la teoría de estimación involucrada, se desarrollarán algunos resultados que llevan a los estimadores máximo verosímiles para $\alpha, \beta, \alpha_\perp$ y $\beta_\perp$. Adicionalmente, se presentará, sin demostración una prueba de hipótesis para indagar sobre las variables de las que depende $\alpha_\perp$, el coeficiente que define el espacio de tendencias comunes.

El desarrollo que lleva a los estimadores máximo verosímiles para $\alpha$ y $\beta$ es una adaptación, para su aplicación a la estimación del MCE, de la exposición de Reinsel y Velu (1998).

### 2.3.1 Regresión de Rango Reducido

El modelo VAR en su forma MCE se puede reexpresar de la siguiente forma:

$$Y = \nabla X_t = \Gamma_0 + \alpha \beta' X_{t-1} - \Gamma_1 \nabla X_{t-1} - \Gamma_2 \nabla X_{t-2} - \cdots - \Gamma_k \nabla X_{t-k+1} + \varepsilon_t = ABX + DW + \varepsilon$$

donde,
- $Y = \nabla X_t$, $\varepsilon = \varepsilon_t$
- $\alpha = A \in \mathbb{R}^{pxr}$ y $\beta' = B \in \mathbb{R}^{rxp}$,
- $D = [\Gamma_1, \Gamma_2, \ldots, \Gamma_k] \in \mathbb{R}^{pkxp}$, y
- $W = [\nabla X'_{t-1}, \nabla X'_{t-2}, \ldots, \nabla X'_{t-k+1}]'$

En lo subsiguiente se considera que Y, $X$ y W se han corregido de forma que tienen esperanza igual a cero. En este caso, nos interesa estimar A y B por lo que podemos *concentrar* el modelo con respecto al coeficiente D mediante dos modelos de regresión auxiliares:

$$Y = FW + \varepsilon_0 \quad y \quad X = GW + \varepsilon_1$$

En donde $W$ está dado, de forma que

$$Y = ABX + DW + \varepsilon \Rightarrow FW + \varepsilon_0 = AB(GW + \varepsilon_1) + DW + \varepsilon \Rightarrow$$
$$\varepsilon_0 = (ABG + D - F)W + AB\varepsilon_1 + \varepsilon$$

Pero como $Cov(\varepsilon_0, W) = 0$ se tiene la siguiente relación:

$$\varepsilon_0 = AB\varepsilon_1 + \varepsilon$$

de donde se puede estimar A y B a partir de una regresión multivariada de rango reducido:

$$R_0 = ABR_1 + \varepsilon$$

donde:
- $R_0 = Y - \hat{F}W = \hat{\varepsilon_0}$
- $R_1 = X - \hat{G}W = \hat{\varepsilon_1}$
- $A \in \mathbb{R}^{pxr}, B \in \mathbb{R}^{rxp}$

En las siguientes secciones se desarrollan los estimadores de mínimos cuadrados y de máxima verosimilitud para $A \in \mathbb{R}^{mxr}, B \in \mathbb{R}^{rxn}$ aún cuando, en nuestro caso, sabemos que $m = n = p$.

### 2.3.2 Estimadores de Mínimos Cuadrados

Considerar el modelo:
$$Y = ABX + \varepsilon \quad \varepsilon \sim N(0, \Sigma_{\varepsilon\varepsilon}) \; iid$$

Los estimadores de rango reducido para A y B tal que $C = AB$ con $rank(C) = r \leq \min(m, n)$, se obtendrán como la aproximación de una matriz de rango completo, a partir de una matriz de rango reducido. Para llegar al correspondiente resultado necesitaremos el teorema de Eckart-Young, pero antes necesitamos el siguiente lema.

**Lema 2.3.2.1**: Sea $A \in \mathbb{R}^{mxm}$ simétrica con eigenvalores $\lambda_1 \geq \lambda_2 \geq \cdots \geq \lambda_m$ y eigenvectores (normalizados) $P_1, P_2, \ldots, P_m$. Entonces el supremo de:

$$\sum_{i=1}^{r} X_i'AX_i = tr(X'AX)$$

sobre los conjuntos $\{X_1, X_2, \ldots, X_r\}$ con $X_i \in \mathbb{R}^m$ tal que $X_i'X_i = 1$ y $X_i'X_j = 0$ con $i \neq j$, es igual a $\sum_{i=1}^{r} \lambda_i$ y se obtiene haciendo $X_i = P_i$ para $i \in \{1,2,\ldots,r\}$.

**Demostración**. Como $A$ simétrica existe $P \in \mathbb{R}^{mxm}$ ortogonal tal que :
$$A = PDP' \quad y \quad D = P'AP$$

En donde $P = [P_1, P_2, \ldots, P_m]$ son los eigenvectores normalizados de $A$. Como $P$ es ortogonal, es de rango completo por lo que cualquier conjunto ortogonal $\{X_1, X_2, \ldots, X_r\}$ se puede expresar utilizando que para toda $X_i \in \mathbb{R}^m$ existe $c_i \in \mathbb{R}^m$ tal que $X_i = Pc_i$. Con esto obtenemos las siguientes relaciones:

$$X_i'X_i = c_i'P'Pc_i = c_i'c_i = 1$$
$$X_i'X_j = c_i'P'Pc_j = c_i'c_j = 0$$

Además, la cantidad a maximizar se puede expresar de la siguiente forma:

$$\sum_{i=r}^{r} X_i'AX_i = \sum_{i=r}^{r} c_i'P'(PDP')Pc_i = \sum_{i=r}^{r} c_i'Dc_i = \sum_{i=1}^{r}\sum_{j=1}^{m} \lambda_j c_{ij}^2 = \sum_{j=1}^{m} \lambda_j \left(\sum_{i=1}^{r} c_{ij}^2\right) = \sum_{j=1}^{m} \lambda_j a_j$$

Observar que:
- $a_j = \sum_{i=1}^{r} c_{ij}^2$
- $c_i'c_i = \sum_{i=1}^{m} c_{ij}^2 = 1 \geq \sum_{i=1}^{r} c_{ij}^2$
- $\sum_{i=1}^{r}\sum_{j=1}^{m} c_{ij}^2 = \sum_{i=1}^{r} c_i'c_i = r$

Entonces, $\sum_{i=r}^{r} X_i'AX_i = \sum_{j=1}^{m} \lambda_j a_j$ es una combinación lineal de los eigenvalores de $A$ en la que los coeficientes $a_j$ suman $r$ pero ninguno es mayor a 1. Como $\lambda_1 \geq \lambda_2 \geq \cdots \geq \lambda_m$ esta expresión se maximiza cuando el coeficiente $a_j = 1$ para $j \in \{1,2,\ldots,r\}$ y $a_j = 0$ para $j \in \{r+1, r+2, \ldots, m\}$. Como además, $c_i'c_i = 1$ y $c_i'c_j = 0$ esto se logra haciendo $c_{ii} = 1$ y $c_{ij} = 0$. Con está elección tenemos que:

1. $X_i = Pc_i = P_i$

2. $\sum_{i=r}^{r} X_i'AX_i = \sum_{j=1}^{m} \lambda_j \left(\sum_{i=1}^{r} c_{ij}^2\right) = \sum_{j=1}^{r} \lambda_j$

*Q.E.D.*

A continuación, se enuncia el teorema Eckart-Young que nos indicará cómo mejor aproximar una matriz de rango completo con una de rango reducido.

**Teorema 2.3.2.1 (Eckart-Young)**: Sea $S \in \mathbb{R}^{mxn}$ una matriz fija tal que $rank(S) = m$. Entonces:

$$arg\min_{P \in RR} tr\big((S-P)(S-P)'\big) = N'NS'$$

En donde $RR = \{Q \in \mathbb{R}^{mxr} : rank(Q) = r \leq m\}$ y $N \in \mathbb{R}^{rxm}$ es tal que las columnas de $N'$ son los $r$ eigenvectores (normalizados) de $S'S$ que corresponden a sus $r$ eigenvalores más grandes.

**Demostración**. Sea $P = MN$ y suponer, sin perdida de generalidad, que $N'$ es ortonormal ($NN' = I_r$) de tal forma que la columna $i$ de $P'$ se puede expresar como una combinación lineal de las columnas de $N'$ en la que los coeficientes provienen de la $i$-esima columna de $M'$. Ahora considerar el criterio a minimizar:

$$\begin{aligned} f(P) &= tr\big((S-P)(S-P)'\big) = tr\big((S-P)'(S-P)\big) = tr\big((S'-P')(S'-P')'\big) \\ &= tr\big((S'-N'M')(S-MN)\big) = tr(S'S - S'MN - N'M'S + N'M'MN) \\ &= tr(S'S) - 2tr(S'MN) + tr(N'M'MN) = tr(S'S) - tr(MNS') + tr(MM') \end{aligned}$$

Para $N$ fija podemos minimizar $f(P)$, que es continua como función de $M$, derivando con respecto a $M$ e igualando a cero:

$$\frac{\delta f(P)}{\delta M} = -2SN' + 2M$$

$$\frac{\delta f(P)}{\delta M} = 0 \Rightarrow M = SN', \quad P = MN = SN'N$$

Sustituyendo en $f(P)$ tenemos que:

$$f(P) = tr\big((S' - P')(S' - P')'\big) = tr\big((S' - N'M')(S' - N'M')'\big)$$
$$= tr\big((S' - N'NS')(S' - N'NS')'\big)$$
$$= tr(S'S) - 2tr(S'SN'N) + tr(N'NS'SN'N)$$
$$= tr(S'S) - 2tr(S'SN'N) + tr(S'SN'N) = tr(S'S) - tr(S'SN'N)$$
$$= tr(S'S) - tr(NS'SN')$$

Entonces para minimizar $f(P)$ con respecto $N$ utilizamos el Lema anterior. De acuerdo a este lema el supremo de $tr((N')'(S'S)N')$, se alcanza cuando $N'$ contiene los eigenvectores (normalizados) de $S'S$ en cual caso el valor mínimo será $\sum_{i=1}^{r} \lambda_i^2$, donde $\lambda_i^2$ es el $i$-esimo eigenvalor más grande de $S'S$.

### *Q.E.D.*

Ahora, para obtener el valor del criterio a minimizar en el teorema utilizamos la descomposición en valores singulares de $S = VDU'$:

$$S = VDU' \quad S'S = UDV'VDU' = U'D^2U$$

En donde,
- $V \in \mathbb{R}^{m \times m}$ es una matriz ortogonal que contiene vectores singulares derechos de $S$ y eigenvectores de $SS'$.
- $U \in \mathbb{R}^{n \times m}$ es una matriz ortogonal que contiene vectores singulares izquierdos de $S$ y eigenvectores de $S'S$.
- $D \in \mathbb{R}^{m \times m}$ es una matriz diagonal con valores singulares de $S$ y raíz cuadrada de los eigenvalores de $SS'$ y $S'S$.

Observar que $N' = [U_{*1}, U_{*2}, \ldots, U_{*r}] = U_{(r)} \in \mathbb{R}^{n \times r}$. Entonces el criterio a minimizar queda:

$$tr\big((S' - P')(S' - P')'\big) = tr\big((UDV' - N'M')(UDV' - N'M')'\big)$$
$$= tr(U(D - U'N'M'V)(D - U'N'M'V)'U')$$
$$= tr\big((D - U'N'M'V)(D - U'N'M'V)'\big)$$

Ahora por el teorema 2.3.2.1, $P$ que minimiza el criterio es tal que $P' = N'M' = N'NS'$ y $M'$ se puede reformular de la siguiente forma:

$$M' = NS' = U'_{(r)}S' = U'_{(r)}UDV' = [I_r \ 0]DV' = [\lambda_1 V_{*1} \ \lambda_2 V_{*2} \ \ldots \ \lambda_r V_{*r}] = D_{(r)}V'_{(r)}$$

En donde,
- $V_{(r)} = [V_{*1}, V_{*2}, \ldots, V_{*r}]$
- $D_{(r)} = diag(\lambda_1, \lambda_2, \ldots, \lambda_r)$

Regresando al criterio a minimizar tenemos que:

$$tr\big((S' - P')(S' - P')'\big) = tr\big((D - U'N'M'V)(D - U'N'M'V)'\big)$$
$$= tr(DD') - 2tr(DV'MNU) + tr(U'N'M'VV'MNU)$$

Pero observar que $U'N'M'V = U'U'_{(r)}D_{(r)}V'_{(r)}V = \begin{bmatrix} I_r \\ 0 \end{bmatrix} D_{(r)} [I_r \quad 0] = \begin{bmatrix} D_{(r)} & 0 \\ 0 & 0 \end{bmatrix}$

Así que,

$$tr\big((S' - P')(S' - P')'\big) = tr(D^2) - tr(D_{(r)}^2) = \sum_{i=r+1}^{m} \lambda_i^2$$

Sabiendo como aproximar una matriz de rango completo con una de rango reducido estamos en condiciones de enunciar el teorema que indica como obtener los estimadores de mínimos cuadrados para A y B.

**Teorema 2.3.2.2**: Suponer que el vector $(m+n)$-dimensional $(Y', X')$ tiene media $\bar{\bar{0}} \in \mathbb{R}^{m+n}$ y matriz de covarianza $\Sigma_{yx} = \Sigma_{xy} = Cov(Y, X)$ y que $\Sigma_{xx} = Var(X)$ es no singular. Entonces, para cualquier matriz simétrica positiva definida $\Gamma$, las matrices $A \in \mathbb{R}^{mxr}$ y $B^{rxn}$, con $r \leq \min(m, n)$, que minimizan:

$$tr\{E[\Gamma^{1/2}(Y - ABX)(Y - ABX)'\Gamma^{1/2}]\}$$

son

$$A = \Sigma_{yx}\Sigma_{xx}^{-1/2}U_{(r)} \quad B = U'_{(r)}\Sigma_{xx}^{-1/2}$$

donde $U_{(r)} = [U_{*1}, U_{*2}, \ldots, U_{*r}]$ y $U_{*i}$ es el eigenvector (normalizado) que corresponde al $i$-esimo eigenvalor más grande, $\lambda_i^2$, de la matriz $\Sigma_{xx}^{-1/2}\Sigma_{xy}\Gamma\Sigma_{yx}\Sigma_{xx}^{-1/2}$.

**Demostración**. Buscamos expresar la cantidad $tr\{E[\Gamma^{1/2}(Y - ABX)(Y - ABX)'\Gamma^{1/2}]\}$ de tal forma que podamos usar el teorema 2.3.2.1. que es para expresiones de la forma $tr((S - P)(S - P)')$.

$$E[\Gamma^{1/2}(Y - ABX)(Y - ABX)'\Gamma^{1/2}]$$
$$= \Gamma^{1/2}(E[YY'] - ABE[XY'] - E[YX']B'A' + ABE[XX']B'A')\Gamma^{1/2}$$
$$= \Gamma^{1/2}(\Sigma_{yy} - AB\Sigma_{xy} - \Sigma_{yx}B'A' + AB\Sigma_{xx}B'A')\Gamma^{1/2}$$
$$= \Gamma^{1/2}\left((\Sigma_{yx}\Sigma_{xx}^{-1/2} - AB\Sigma_{xx}^{1/2})(\Sigma_{yx}\Sigma_{xx}^{-1/2} - AB\Sigma_{xx}^{1/2})' - \Sigma_{yx}\Sigma_{xx}^{-1}\Sigma_{xy} + \Sigma_{yy}\right)\Gamma^{1/2}$$

Por lo tanto

$$tr\{E[\Gamma^{1/2}(Y - ABX)(Y - ABX)'\Gamma^{1/2}]\}$$
$$= tr\left\{\Gamma^{1/2}(\Sigma_{yx}\Sigma_{xx}^{-1/2} - AB\Sigma_{xx}^{1/2})(\Sigma_{yx}\Sigma_{xx}^{-1/2} - AB\Sigma_{xx}^{1/2})'\Gamma^{1/2}\right\}$$
$$+ tr\{\Gamma^{1/2}(\Sigma_{yy} - \Sigma_{yx}\Sigma_{xx}^{-1}\Sigma_{xy})\Gamma^{1/2}\}$$

Para minimizar esta expresión con respecto a A y B hay que minimizar

$$tr((S - P)(S - P)') = tr\left\{\Gamma^{1/2}(\Sigma_{yx}\Sigma_{xx}^{-1/2} - AB\Sigma_{xx}^{1/2})(\Sigma_{yx}\Sigma_{xx}^{-1/2} - AB\Sigma_{xx}^{1/2})'\Gamma^{1/2}\right\}$$

con $S = \Gamma^{1/2}\Sigma_{yx}\Sigma_{xx}^{-1/2}$ y $P = \Gamma^{1/2} AB\Sigma_{xx}^{1/2}$. Entonces por el teorema 2.3.2.1.

$$P = MN = SN'N = \Gamma^{1/2}\Sigma_{yx}\Sigma_{xx}^{-1/2}U_{(r)}U'_{(r)} = \Gamma^{1/2} AB\Sigma_{xx}^{1/2} \Rightarrow AB$$

$$= \left(\Sigma_{yx}\Sigma_{xx}^{-1/2}U_{(r)}\right)\left(U'_{(r)}\Sigma_{xx}^{-1/2}\right)$$

con $A = \Sigma_{yx}\Sigma_{xx}^{-1/2}U_{(r)}$ y $B = U'_{(r)}\Sigma_{xx}^{-1/2}$.

***Q.E.D.***

En cuanto al criterio a minimizar habíamos visto que $tr((S - P)(S - P)') = \sum_{i=r+1}^{m}\lambda_i^2 = \sum_{i=1}^{m}\lambda_i^2 - \sum_{i=1}^{r}\lambda_i^2$. Falta considerar el segundo sumando:

$$tr\{\Gamma^{1/2}(\Sigma_{yy} - \Sigma_{yx}\Sigma_{xx}^{-1}\Sigma_{xy})\Gamma^{1/2}\} = tr\{\Gamma^{1/2}\Sigma_{yy}\Gamma^{1/2}\} - tr\{\Gamma^{1/2}\Sigma_{yx}\Sigma_{xx}^{-1}\Sigma_{xy}\Gamma^{1/2}\}$$
$$= tr\{\Gamma\Sigma_{yy}\} - tr\{SS'\} = tr\{\Gamma\Sigma_{yy}\} - \sum_{i=1}^{m}\lambda_i^2$$

Por lo tanto si $A = \Sigma_{yx}\Sigma_{xx}^{-1/2}U_{(r)}$ y $B = U'_{(r)}\Sigma_{xx}^{-1/2}$ entonces $tr\{E[\Gamma^{1/2}(Y - ABX)(Y - ABX)'\Gamma^{1/2}]\} = tr\{\Gamma\Sigma_{yy}\} - \sum_{i=1}^{r}\lambda_i^2$.

El teorema 2.3.2.2 muestra que considerando a $X$ como una variable aleatoria las matrices A y B que minimizan la esperanza de la suma de errores cuadrados reescalados (por la matriz $\Gamma$) son $A = \Sigma_{yx}\Sigma_{xx}^{-1/2}U_{(r)}$ y $B = U'_{(r)}\Sigma_{xx}^{-1/2}$. En la siguiente sección se mostrará que con la elección $\Gamma = \Sigma_{\varepsilon\varepsilon}^{-1}$ o $\Gamma = \Sigma_{yy}^{-1}$, y sustituyendo las matrices de covarianza, $\Sigma_{yx}$, $\Sigma_{xx}$ y $\Sigma_{yy}$ por sus estimadores muestrales, obtenemos los estimadores de máxima verosimilitud.

### 2.3.3 Estimadores de Máxima Verosimilitud

Para demostrar que los estimadores máximo verosímiles son de cierta forma necesitaremos un teorema de separación incluido en Rao (1979). Se incluye el teorema sin demostración.

**Teorema 2.3.3.1 (Teorema de Separación).** Sea $S, P \in \mathbb{R}^{mxn}$ con $rank(S) = m$ y $rank(P) = r \leq m$. Entonces:

$$* \lambda_i(S - P) \geq \lambda_{r+i}(S) \quad \forall i \in \{1,2,\dots,m\}$$

donde:
- $\lambda_i(S)$ denota el $i$-esimo valor singular más grande de $S = VDU'$
- $\lambda_{r+i}(S)$ se define como 0 para $r + i > m$.

La igualdad se obtiene si y solo si $P = V_{(r)}D_{(r)}U'_{(r)}$

Obtengamos la log-verosimilitud del modelo:

$$Y = CX + \varepsilon = ABX + \varepsilon \quad \varepsilon \sim N(0, \Sigma_{\varepsilon\varepsilon}) \text{ iid}$$

donde
- $X \in \mathbb{R}^{nxT}, Y \in \mathbb{R}^{mxT}$ las matrices de realizaciones observadas.
- $\varepsilon \in \mathbb{R}^{mxT}$ una matriz de realizaciones no observables
- $C \in \mathbb{R}^{mxn}, A \in \mathbb{R}^{mxr}$ y $B \in \mathbb{R}^{rxn}$,
- $\Sigma_{\varepsilon\varepsilon} \in \mathbb{R}^{mxm}$

En este caso tenemos que la función de densidad de una $\varepsilon_{*i} = Y_{*i} - CX_{*i}$ es:

$$f_\varepsilon(\varepsilon_{*i}) = (2\pi)^{-m/2}|\Sigma_{\varepsilon\varepsilon}|^{-1/2} exp\left\{-\frac{1}{2}\varepsilon'_{*i}\Sigma_{\varepsilon\varepsilon}^{-1}\varepsilon_{*i}\right\}$$

Por lo que la función de verosimilitud es:

$$\mathcal{L}(C,\Sigma_{\varepsilon\varepsilon};\varepsilon) = (2\pi)^{-mT/2}|\Sigma_{\varepsilon\varepsilon}|^{-T/2} exp\left\{-\frac{1}{2}\sum_{i=1}^{T}\varepsilon'_{*i}\Sigma_{\varepsilon\varepsilon}^{-1}\varepsilon_{*i}\right\}$$

$$= (2\pi)^{-mT/2}|\Sigma_{\varepsilon\varepsilon}|^{-T/2} exp\left\{-\frac{1}{2}tr(\varepsilon'\Sigma_{\varepsilon\varepsilon}^{-1}\varepsilon)\right\}$$

$$= (2\pi)^{-mT/2}|\Sigma_{\varepsilon\varepsilon}|^{-T/2} exp\left\{-\frac{1}{2}tr(\Sigma_{\varepsilon\varepsilon}^{-1}\varepsilon\varepsilon')\right\}$$

Tomando logaritmo natural y omitiendo términos que no dependan de $C$ o $\Sigma_{\varepsilon\varepsilon}$ tenemos que:

$$ln\mathcal{L}(C,\Sigma_{\varepsilon\varepsilon};\varepsilon) \propto \left(\frac{T}{2}\right)[\ln(|\Sigma_{\varepsilon\varepsilon}^{-1}|) - tr(\Sigma_{\varepsilon\varepsilon}^{-1}W)]$$

donde $W = \frac{1}{T}\varepsilon\varepsilon' = \frac{1}{T}(Y-CX)(Y-CX)' = \frac{1}{T}(Y-ABX)(Y-ABX)'$. Derivamos para maximizar con respecto a $\Sigma_{\varepsilon\varepsilon}^{-1}$:

$$\frac{\delta ln\mathcal{L}}{\delta\Sigma_{\varepsilon\varepsilon}^{-1}} = \left(\frac{T}{2}\right)[\Sigma_{\varepsilon\varepsilon} - W]$$

por lo tanto $\hat{\Sigma}_{\varepsilon\varepsilon} = W$. Concentramos la log-verosimilitud con respecto a $\Sigma_{\varepsilon\varepsilon}^{-1}$ sustituyendo $\hat{\Sigma}_{\varepsilon\varepsilon} = W$:

$$ln\mathcal{L}(C;\varepsilon) \propto \left(\frac{T}{2}\right)[\ln(|W^{-1}|) - tr(W^{-1}W)] = -\left(\frac{T}{2}\right)[\ln(|W|) + m]$$

Maximizar esta expresión es equivalente a maximizar $\alpha|W|$ con $\alpha > 0$. En particular podemos escoger $\alpha = |\tilde{\Sigma}_{\varepsilon\varepsilon}^{-1}| = |(1/T)(Y-\tilde{C}X)(Y-\tilde{C}X)'|$, con $\tilde{C} = YX'(XX')^{-1}$, donde $\tilde{\Sigma}_{\varepsilon\varepsilon}^{-1}$ y $\tilde{C}$ son los estimadores de mínimos cuadrados correspondientes al modelo de rango completo. Al escoger $\alpha = |\tilde{\Sigma}_{\varepsilon\varepsilon}^{-1}|$ se facilitará la maximización de $ln\mathcal{L}(C;\varepsilon)$.

Entonces, hay que minimizar $\alpha|W| = |\tilde{\Sigma}_{\varepsilon\varepsilon}^{-1}||W| = |\tilde{\Sigma}_{\varepsilon\varepsilon}^{-1}W|$. Podemos expresar W en términos de $\tilde{\Sigma}_{\varepsilon\varepsilon}$ y $\hat{\Sigma}_{xx} = \frac{1}{T}XX'$:

$$W = \frac{1}{T}\varepsilon\varepsilon' = \frac{1}{T}(Y - ABX)(Y - ABX)' = \frac{1}{T}(Y - \tilde{C}X + \tilde{C}X + ABX)(Y - \tilde{C}X + \tilde{C}X + ABX)'$$

$$= \frac{1}{T}\left((Y - \tilde{C}X) + (\tilde{C} - AB)X\right)\left((Y - \tilde{C}X) + X'(\tilde{C} - AB)\right)'$$

$$= \frac{1}{T}\left((Y - \tilde{C}X)(Y - \tilde{C}X)' + (\tilde{C} - AB)X(Y - \tilde{C}X)' + (Y - \tilde{C}X)X'(\tilde{C} - AB)'\right.$$

$$\left. + (\tilde{C} - AB)XX'(\tilde{C} - AB)'\right)$$

Pero $(\tilde{C} - AB)X(Y - \tilde{C}X)' = (YX'(XX')^{-1} - AB)X(Y - YX'(XX')^{-1}X)' =$
$(YX'(X')^{-1}X^{-1}X - ABX)(Y - YX'(X')^{-1}X^{-1}X)' = (Y - ABX)(Y - Y)' = 0$
Así que

$$W = \frac{1}{T}\left((Y - \tilde{C}X)(Y - \tilde{C}X)' + (\tilde{C} - AB)XX'(\tilde{C} - AB)'\right) = \tilde{\Sigma}_{\varepsilon\varepsilon} + (\tilde{C} - AB)\hat{\Sigma}_{xx}(\tilde{C} - AB)'$$

Con esto el criterio a minimizar para obtener el estimador máximo verosímil de $A$ y $B$ es:
$$\left|\tilde{\Sigma}_{\varepsilon\varepsilon}^{-1}W\right| = \left|\tilde{\Sigma}_{\varepsilon\varepsilon}^{-1}\left(\tilde{\Sigma}_{\varepsilon\varepsilon} + (\tilde{C} - AB)\hat{\Sigma}_{xx}(\tilde{C} - AB)'\right)\right| = |I_m + \tilde{\Sigma}_{\varepsilon\varepsilon}^{-1}(\tilde{C} - AB)\hat{\Sigma}_{xx}(\tilde{C} - AB)'|$$

El determinante de una matriz es igual al producto de sus eigenvalores. Además si $Q \neq I$ y $\lambda^2$ es eigenvalor de $Q$ entonces $1 + \lambda^2$ es eigenvalor de $I + Q$ por lo que:

$$\left|\tilde{\Sigma}_{\varepsilon\varepsilon}^{-1}W\right| = \left|I_m + \tilde{\Sigma}_{\varepsilon\varepsilon}^{-1}(\tilde{C} - AB)\hat{\Sigma}_{xx}(\tilde{C} - AB)'\right| = \prod_{i=1}^{m}(1 + \lambda_i^2)$$

donde $\lambda_i^2$ es el $i$-esimo eigenvalor más grande de $\tilde{\Sigma}_{\varepsilon\varepsilon}^{-1}(\tilde{C} - AB)\hat{\Sigma}_{xx}(\tilde{C} - AB)'$. Los eigenvalores de $R^{-1}Q$ también son eigenvalores de $R^{-1/2}QR^{-1/2}$ siempre y cuando R sea simétrica positiva definida así que $\lambda_i^2$ es el $i$-esimo eigenvalor más grande de $\tilde{\Sigma}_{\varepsilon\varepsilon}^{-1/2}(\tilde{C} -$ $AB)\hat{\Sigma}_{xx}(\tilde{C} - AB)'\tilde{\Sigma}_{\varepsilon\varepsilon}^{-1/2}$. Por lo tanto minimizar $|\tilde{\Sigma}_{\varepsilon\varepsilon}^{-1}W|$ corresponde a minimizar $\prod_{i=1}^{m}(1 + \lambda_i^2)$ donde $\lambda_i^2$ son los eigenvalores de :

$$\tilde{\Sigma}_{\varepsilon\varepsilon}^{-1/2}(\tilde{C} - AB)\hat{\Sigma}_{xx}(\tilde{C} - AB)'\tilde{\Sigma}_{\varepsilon\varepsilon}^{-1/2} = (S - P)(S - P)'$$

con

$$S = \tilde{\Sigma}_{\varepsilon\varepsilon}^{-1/2}\tilde{C}\hat{\Sigma}_{xx}^{1/2} \quad P = \tilde{\Sigma}_{\varepsilon\varepsilon}^{-1/2}AB\hat{\Sigma}_{xx}^{1/2}$$

Observar que $S$ está fija y $P$ hay que determinarla de tal forma que los eigenvalores, $\lambda_i^2$, de $= (S - P)(S - P)'$ se minimicen simultáneamente, lo que equivale a que se minimicen los valores singulares, $\lambda_i$, de $S - P$. Ahora por el teorema 2.3.3.1 todos los valores singulares de $S - P$ se minimizan simultáneamente si y solo sí $P = V_{(r)}D_{(r)}U'_{(r)}$ sonde $S = VDU'$ es la descomposición en valores singulares de $S$. Recordando que $P = MN$ con $N = U'_{(r)}$ y $M = V_{(r)}D_{(r)} = SU_{(r)}$ tenemos que:

$$P = SU_{(r)}U'_{(r)} = \left(\tilde{\Sigma}_{\varepsilon\varepsilon}^{-1/2}\tilde{C}\hat{\Sigma}_{xx}^{1/2}\right)U_{(r)}U'_{(r)} = \tilde{\Sigma}_{\varepsilon\varepsilon}^{-1/2}AB\hat{\Sigma}_{xx}^{1/2} \Rightarrow AB = \left(\tilde{C}\hat{\Sigma}_{xx}^{1/2}U_{(r)}\right)\left(U'_{(r)}\hat{\Sigma}_{xx}^{-1/2}\right)$$

con A $= \tilde{C}\hat{\Sigma}_{xx}^{1/2}U_{(r)} = YX'(XX')^{-1}\hat{\Sigma}_{xx}^{1/2}U_{(r)} = \hat{\Sigma}_{yx}\hat{\Sigma}_{xx}^{-1/2}U_{(r)}$ y B $= U'_{(r)}\hat{\Sigma}_{xx}^{-1/2}$.

La única diferencia con los estimadores del teorema 2.3.2.2 es que en ese caso se definió $S = \Gamma^{1/2}\Sigma_{yx}\Sigma_{xx}^{-1/2}$ y en este caso tenemos que $S = \tilde{\Sigma}_{\varepsilon\varepsilon}^{-1/2}\tilde{C}\hat{\Sigma}_{xx}^{1/2} = \tilde{\Sigma}_{\varepsilon\varepsilon}^{-1/2}\hat{\Sigma}_{yx}\hat{\Sigma}_{xx}^{-1/2}$. Es decir que si se selecciona $\Gamma = \tilde{\Sigma}_{\varepsilon\varepsilon}^{-1/2}$ y se estiman las matrices de covarianza con las covarianzas muestrales se obtienen estimadores máximo verosímiles. Resulta que con la selección $\Gamma = \hat{\Sigma}_{yy}^{-1} = \frac{1}{T}YY'$ también se obtienen estimadores máximo verosímiles $A_*$ y $B_*$ ya que, aunque $A_* \neq A$ y $B_* \neq B$ resulta que $A_*B_* = AB$. Para ver esto llamemos a $U_{1(r)}$ a la matriz de $r$ eigenvectores de $S_1'S_1$ donde $S_1 = \tilde{\Sigma}_{\varepsilon\varepsilon}^{-1/2}\hat{\Sigma}_{yx}\hat{\Sigma}_{xx}^{-1/2}$ y $U_{2(r)}$ la matriz de $r$ eigenvectores de $S_2'S_2$ donde $S_2 = \hat{\Sigma}_{yy}^{-1/2}\hat{\Sigma}_{yx}\hat{\Sigma}_{xx}^{-1/2}$ . Entonces, tenemos que

$$AB = \hat{\Sigma}_{yx}\hat{\Sigma}_{xx}^{-1/2}U_{1(r)}U'_{1(r)}\hat{\Sigma}_{xx}^{-1/2} \quad y \quad A_*B_* = \hat{\Sigma}_{yx}\hat{\Sigma}_{xx}^{-1/2}U_{2(r)}U'_{2(r)}\hat{\Sigma}_{xx}^{-1/2}$$

Es claro que para que $A_*B_* = AB$, debe ser cierto que $U_{1(r)}U'_{1(r)} = U_{2(r)}U'_{2(r)}$. Para ver que esto es cierto basta ver que $S'_1S_1$ y $S'_2S_2$ tienen el mismo espacio nulo, lo que significa que sus eigenvectores ocupan el mismo subespacio de $\mathbb{R}^n$. Como además, las columnas de $U_{1(r)}$ y $U_{2(r)}$ son ortonormales se puede realizar la siguiente descomposición de la base canónica:

$$U_{1(r)}U'_{1(r)} = U_{2(r)}U'_{2(r)} = I_n - RR'$$

donde las columnas de $R$ son $n-r$ vectores ortonormales que conforman una base para el espacio nulo de $S'_1S_1$ y $S'_2S_2$.

En resumen, para estimar las matrices $A$ y $B$ hay que resolver el problema de eigenvalores:

$$S_2 = \hat{\Sigma}_{yy}^{-1/2}\hat{\Sigma}_{yx}\hat{\Sigma}_{xx}^{-1/2}$$

$$(S'_2S_2)x = \left(\hat{\Sigma}_{xx}^{-1/2}\hat{\Sigma}_{xy}\hat{\Sigma}_{yy}^{-1}\hat{\Sigma}_{yx}\hat{\Sigma}_{xx}^{-1/2}\right)x = \lambda^2 x$$

Como $\hat{\Sigma}_{xx}$ es simétrica positiva definida existe $y \in \mathbb{R}^m$ tal que $x = \hat{\Sigma}_{xx}^{1/2}y$. Sustituyendo en la ecuación anterior y multiplicando por la izquierda por $\hat{\Sigma}_{xx}^{1/2}$ llegamos al siguiente problema de eigenvalores generalizado:

$$\hat{\Sigma}_{xx}^{1/2}\left(\hat{\Sigma}_{xx}^{-1/2}\hat{\Sigma}_{xy}\hat{\Sigma}_{yy}^{-1}\hat{\Sigma}_{yx}\hat{\Sigma}_{xx}^{-1/2}\right)(\hat{\Sigma}_{xx}^{1/2}y) = \lambda^2\hat{\Sigma}_{xx}^{1/2}\left(\hat{\Sigma}_{xx}^{1/2}y\right) \Rightarrow \left(\hat{\Sigma}_{xy}\hat{\Sigma}_{yy}^{-1}\hat{\Sigma}_{yx}\right)y = \lambda^2\hat{\Sigma}_{xx}y$$

La solución para $A$ y $B$ en términos de los eigenvalores del problema generalizado es más sencilla:

$$A = \hat{\Sigma}_{yx}\hat{\Sigma}_{xx}^{-1/2}U_{(r)} = \hat{\Sigma}_{yx}W_{(r)} \quad y \quad B' = \hat{\Sigma}_{xx}^{-1/2}U_{(r)} = W_{(r)}$$

Con $W_{(r)} = \hat{\Sigma}_{xx}^{-1/2}U_{(r)} = [\hat{\Sigma}_{xx}^{-1/2}U_{*1}, \hat{\Sigma}_{xx}^{-1/2}U_{*2}, \ldots, \hat{\Sigma}_{xx}^{-1/2}U_{*r}] = [W_{*1}, W_{*2}, \ldots, W_{*r}]$ donde $W_{*i}$ es el eigenvector del problema generalizado que corresponde al $i$-ésimo eigenvalor más grande.

En términos del problema del análisis de cointegración y la estimación de α y β en el MCE:

$$\nabla X_t = \Gamma_0 + \alpha\beta'X_{t-1} - \Gamma_1\nabla X_{t-1} - \Gamma_2\nabla X_{t-2} - \cdots - \Gamma_k\nabla X_{t-k+1} + \varepsilon_t$$

Hay que seguir los siguientes pasos:

1. Calcular $\hat{\varepsilon}_0 = R_0$ del modelo $\nabla X_t = a_0 + a_1\nabla X_{t-1}a_2\nabla X_{t-2} + \cdots + a_k\nabla X_{t-k+1} + \varepsilon_0$
2. Calcular $\hat{\varepsilon}_1 = R_1$ del modelo $X_{t-1} = b_0 + b_1\nabla X_{t-1}b_2\nabla X_{t-2} + \cdots + b_k\nabla X_{t-k+1} + \varepsilon_1$
3. Calcular $S_{10} = \widehat{Cov}(R_1, R_0) = \frac{1}{T}R_1R_0$, $S_{00} = \widehat{Cov}(R_0, R_0) = \frac{1}{T}R_0R_0$ y $S_{11} =$ $\widehat{Cov}(R_1, R_1) = \frac{1}{T}R_1R_1$
4. Resolver el problema de eigenvectores-eigenvalores:

$$(S_{10}S_{00}^{-1}S_{01})y = \lambda^2 S_{11}y$$

5. Calcular $\hat{\alpha}$ y $\hat{\beta}$:

$$\hat{\alpha} = S_{01}W_{(r)} \quad y \quad \hat{\beta} = W_{(r)}$$

con $W_{(r)} = [W_{*1}, W_{*2}, \ldots, W_{*r}]$ donde $W_{*i}$ es el eigenvector del problema generalizado que corresponde al $i$-ésimo eigenvalor más grande.

Para encontrar los estimadores de $\alpha_\perp$ y $\beta_\perp$ considerar el problema de eigenvectores-eigenvalores anterior (primal) y el problema dual asociado:

$$(S_{10}S_{00}^{-1}S_{01})x = \lambda^2 S_{11}x$$
$$(S_{01}S_{11}^{-1}S_{10})y = \delta^2 S_{00}y$$

Partiendo del la primera ecuación, multiplicando por la derecha por $S_{01}S_{11}^{-1}$ obtenemos:

$$S_{01}S_{11}^{-1}(S_{10}S_{00}^{-1}S_{01})x = \lambda^2 S_{01}x$$

Sustituyendo $w = S_{01}x$ se obtiene:
$$(S_{01}S_{11}^{-1}S_{10})S_{00}^{-1}w = \lambda^2 w$$

Ahora como $S_{00}$ es simétrica positiva definida, existe $y \in \mathbb{R}^p$ tal que $w = S_{00}y$ con lo que llegamos al problema de eigenvectores-eigenvalores dual:

$$(S_{01}S_{11}^{-1}S_{10})y = \lambda^2 S_{00}y$$

Notar que $\lambda^2 = \delta^2$ y que $w = S_{01}x = S_{00}y$ por lo que la relación entre los eigenvectores de los dos problemas es:

$$y = S_{00}^{-1}S_{01}x$$

Donde x es eigenvector del problema primal y y del dual. De forma análoga se obtiene que:

$$x = S_{11}^{-1}S_{10}y$$

Sea que $W = [W_{*1}, W_{*2}, \ldots, W_{*p}]$, $W_{(r)} = [W_{*1}, W_{*2}, \ldots, W_{*r}]$ y $W_{(k)} = [W_{*r+1}, W_{*r+2}, \ldots, W_{*p}]$ donde $k = p - r$ y $W_{*i}$ es el eigenvector del problema primal asociado al $i$-ésimo eigenvalor más grande. Análogamente, sea que $Z = [Z_{*1}, Z_{*2}, \ldots, Z_{*p}]$, $Z_{(r)} = [Z_{*1}, Z_{*2}, \ldots, Z_{*r}]$ y $Z_{(k)} = [Z_{*r+1}, Z_{*r+2}, \ldots, Z_{*p}]$ donde $Z_{*i}$ es el eigenvector del problema dual asociado al $i$-ésimo eigenvalor más grande. Entonces tenemos las siguientes relaciones:

$$Z = S_{00}^{-1}S_{01}W \quad y \quad W = S_{11}^{-1}S_{10}Z$$

Recordar que buscamos $\hat{\alpha}_\perp$ y $\hat{\beta}_\perp$ tal que:

$$\hat{\alpha}_\perp' \hat{\alpha} = \hat{\alpha}_\perp' S_{01} W_{(r)} = 0 \quad \hat{\beta}_\perp' \hat{\beta} = \hat{\beta}_\perp' W_{(r)}$$

Ahora los eigenvectores de un problema generalizado de la forma

$$Lx = \lambda^2 Mx$$

donde $M$ es positiva definida, son conjugados con respecto a $M$: es decir que si $x_i$ y $x_j$ son dos eigenvectors distintos se cumple que $x_i' M x_j = 0$. Esto significa que:
$$W_{(k)}' S_{11} W_{(r)} = 0 \in \mathbb{R}^{k \times r}$$
Usando la ecuación $W = S_{11}^{-1}S_{10}Z$, tenemos que :

$$W_{(k)}' S_{11} W_{(r)} = (Z_{(k)}' S_{01} S_{11}^{-1}) S_{11} W_{(r)} = Z_{(k)}' S_{01} W_{(r)} = 0$$

De aquí se puede ver que:

$$\widehat{\alpha}_\perp = Z_{(k)} \quad y \quad \widehat{\beta}_\perp = S_{10}Z_{(k)}$$

Para obtener intervalos de confianza y pruebas de hipótesis para $\widehat{\alpha}$, $\widehat{\beta}$, $\widehat{\alpha}_\perp$ y $\widehat{\beta}_\perp$ es necesario derivar la distribución asintótica de estos estimadores. Para una derivación de la distribución de los estimadores máximo verosímiles se refiere al lector a la exposición de Reinsel y Velu (1998).

### 2.3.4 Otros Resultados

En esta sección se enuncia, sin demostración, un resultado complementario que se usará más adelante, en la aplicación de análisis de cointegración al mercado de deuda de Norteamérica. El siguiente teorema y su demostración se pueden encontrar en Gonzalo y Granger (1995).

**Teorema 2.3.4.1**. Sea $G \in \mathbb{R}^{pxm}$ y $\theta \in \mathbb{R}^{mxk}$ con $k = p - r$ y $k \leq m < p$. Bajo la hipótesis $\mathcal{H}0: \alpha_\perp = G\theta$ se puede encontrar el estimador de máxima verosimilitud para $\alpha_\perp$ de la siguiente forma:

1. Resolver el siguiente problema eigenvalores-eigenvectores generalizado:

$$(G'S_{01}S_{11}^{-1}S_{10}G)y = \lambda^2(G'S_{00}G)y$$

    obteniendo los eigenvalores $\lambda^2_{\mathcal{H}0.1} > \lambda^2_{\mathcal{H}0.2} > \cdots > \lambda^2_{\mathcal{H}0.m}$ y la matriz de eigenvectores normalizados $M = [M_{*1}, M_{*2}, \ldots, M_{*m}]$.

2. Seleccionar

$$\widehat{\theta} = \left[M_{*((m+1)-(p-r))}, \ldots, M_{*m}\right] \quad y \quad \widehat{\alpha}_\perp = G\widehat{\theta}$$

3. La función de verosimilitud maximizada queda:

$$\mathcal{L}_{max}(\mathcal{H}0) \propto |S_{00.1}|\left(\prod_{i=r+1}^{p}\left(1 - \lambda^2_{\mathcal{H}0.(i+m-p)}\right)\right)^{-1}$$

4. Por lo que el estadístico de la prueba de razón de verosimilitudes para la hipótesis $\mathcal{H}0$ versus la hipótesis $\mathcal{H}1: \alpha_\perp = G_*\theta_* \neq G\theta$ con $G_* \in \mathbb{R}^{pxp}$ y $\theta_* \in \mathbb{R}^{pxk}$ es:

$$-2ln(\mathcal{H}0/\mathcal{H}1) = -T\sum_{i=r+1}^{p} ln\{(1 - \lambda^2_{\mathcal{H}0.(i+m-p)})/(1 - \lambda_i^2)\} \sim \chi^2_{(p-r)(p-m)}$$

Está prueba se utilizará para determinar sí $\alpha_\perp$ no depende (asumiendo que cualquier dependencia es lineal) de alguno de los componentes en $X_t$. En este trabajo $G_* = I_p$, $\theta_* = \theta = \widehat{\alpha}_\perp$ y G se conforma de m vectores canónicos distintos, de tal forma que, aquellos vectores canónicos excluidos corresponden a los componentes que se piensa no determinan el valor de $\alpha_\perp$.

## 3  Análisis Exploratorio

A continuación se muestran las tasas de interés de los tres países.

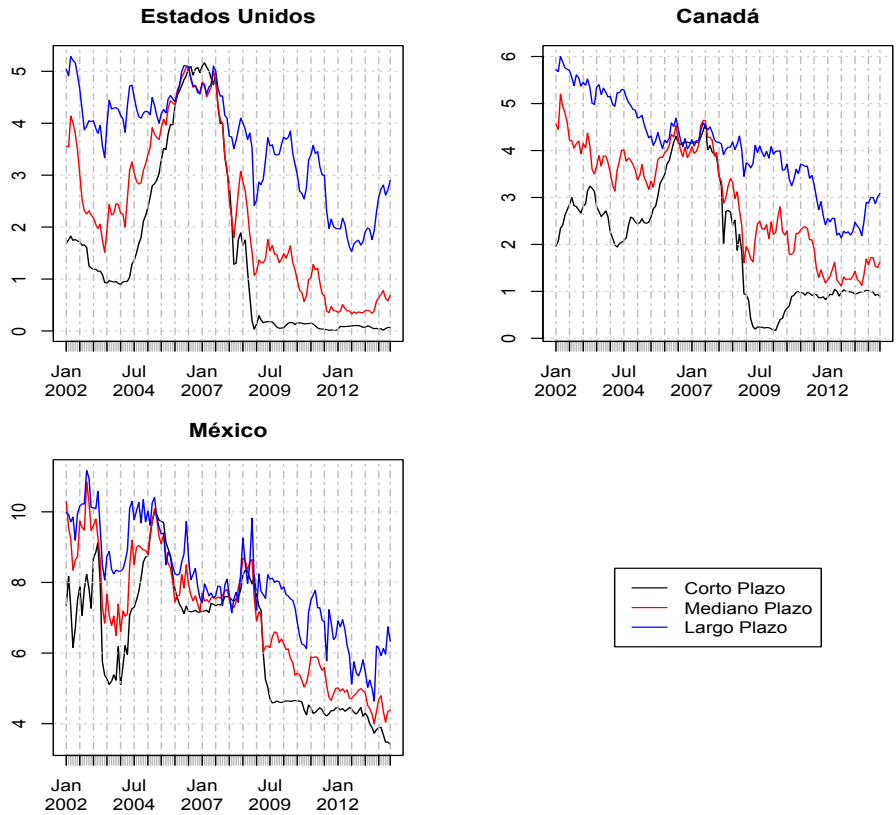

Se puede apreciar que existe cierta evidencia gráfica que las 3 tasas de cada país están cointegradas. A continuación se grafican las tasas de interés agrupándolas por su plazo a vencimiento.

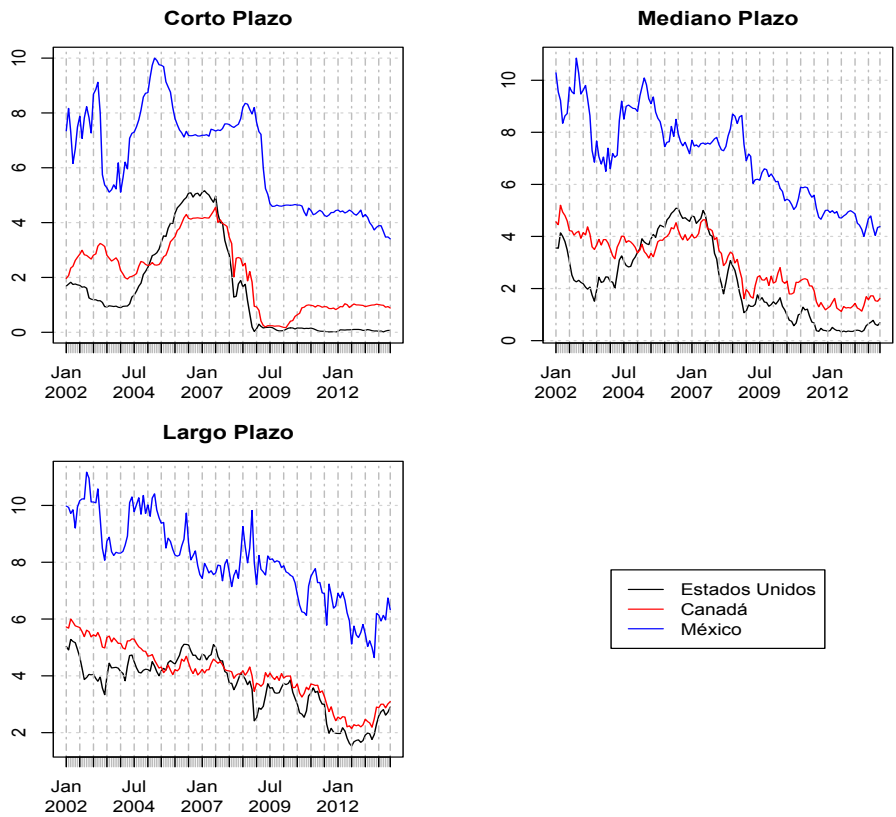

Observando las series de tiempo agrupadas por plazo se nota cierta evidencia gráfica de cointegración especialmente entre las tasas de Estados Unidos y Canadá. Para poder determinar el orden de integración de las series se realiza un análisis de desviaciones estándar de las series y sus diferencias conforme al enfoque de Box y Jenkins. También se realizan pruebas de hipótesis Dickey-Fuller para corroborar la existencia de una raíz unitaria en el polinomio de rezagos autorregresivo.

Para referirnos a los componentes usaremos las siguiente notación:

- Las variables $E$, $C$ y $M$ se refieren a Estados Unidos, Canadá y México
- Los subíndices $CP$, $MP$ y $LP$ se refieren al corto, mediano y largo plazo
- $\nabla_s^d(x_t)$, indica que se toman d diferencias, cada una con rezago de s periodos.

| $x_t$ | Serie original $\sigma(x_t)$ | Primera transformación (optima) | | | Segunda transformación (óptima*) | | |
|---|---|---|---|---|---|---|---|
| | | $s_1$ | $d_1$ | $\sigma\left(\nabla_{s1}^{d1}(x_t)\right)$ | $s_2$ | $d_2$ | $\sigma\left(\nabla_{s2}^{d2}\left(\nabla_{s1}^{d1}(x_t)\right)\right)$ |
| $E_{CP}$ | 1.71 | 1 | 1 | 0.18 | 1 | 1 | 0.19 |
| $E_{MP}$ | 1.50 | 1 | 1 | 0.23 | 1 | 1 | 0.26 |
| $E_{LP}$ | 1.00 | 1 | 1 | 0.23 | 1 | 1 | 0.29 |
| $C_{cp}$ | 1.28 | 1 | 1 | 0.20 | 6 | 1 | 0.25 |
| $C_{Mp}$ | 1.13 | 1 | 1 | 0.23 | 9 | 1 | 0.30 |
| $C_{Lp}$ | 0.99 | 1 | 1 | 0.17 | 9 | 1 | 0.21 |
| $M_{CP}$ | 1.81 | 1 | 1 | 0.41 | 8 | 1 | 0.53 |
| $M_{MP}$ | 1.71 | 1 | 1 | 0.41 | 3 | 1 | 0.55 |
| $M_{LP}$ | 1.46 | 1 | 1 | 0.52 | 4 | 1 | 0.65 |

*Las transformaciones son óptimas, con respecto a $s$ y $d$, en el sentido que minimizan la desviación estándar de la serie

Siguiendo la metodología propuesta por Guerrero (2003) determinamos el orden de integración de la serie observando la desviación estándar de la serie transformada, como función del número de diferencias, y del rezago de éstas. Para cada serie, la desviación estándar se minimiza tomando una diferencia con rezago de un periodo. Esto parece indicar que las series son integradas de orden uno. Confirmamos aplicando pruebas Dickey-Fuller.

| | AIC de AR de orden k | | | | Número de rezagos óptimo | Estadístico Dickey-Fuller Aumentado |
|---|---|---|---|---|---|---|
| | K=0 | K=1 | K=2 | K=3 | | |
| $E_{CP}$ | 565.6 | -73.5 | -104.6 | -102.9 | 2 | -0.89 |
| $E_{MP}$ | 529.3 | -9.4 | -27.4 | -25.4 | 2 | -1.28 |
| $E_{LP}$ | 412.6 | -5.2 | -12.6 | -13.5 | 3 | -1.97 |
| $C_{cp}$ | 482.4 | -50.2 | -48.3 | -51.9 | 3 | -1.14 |
| $C_{Mp}$ | 446.6 | -3.6 | -1.7 | 0.2 | 1 | -1.55 |
| $C_{Lp}$ | 408.5 | -95.0 | -93.6 | -94.4 | 1 | -1.58 |
| $M_{CP}$ | 582.3 | 160.1 | 157.7 | 159.2 | 2 | -1.12 |
| $M_{MP}$ | 565.8 | 162.2 | 164.2 | 166.0 | 1 | -1.13 |
| $M_{LP}$ | 520.0 | 222.1 | 222.0 | 214.9 | 3 | -1.28 |

La tabla muestra el estadístico de prueba Dickey-Fuller Aumentado para cada serie, tomando en cuenta el número de términos autorregresivos óptimo, según el Criterio de Información de Akaike (AIC). Únicamente se tomaron en cuenta rezagos hasta de orden tres tomando en cuenta que se trata de series de tiempo financieras con periodicidad mensual. Los valores críticos para 144 observaciones son -3.46 (1%), -2.88 (5%) y -2.57%, por lo que en ningún caso se rechaza la hipótesis nula de que existe una raíz unitaria.

## 4 Análisis de Cointegración

Se realiza el análisis de cointegración sobre grupos de tasas de interés distintos:
1. Se obtienen los componentes permanentes de cada país y luego se realiza un análisis de cointegración de todos los componentes permanentes.
2. Se realiza el análisis de cointegración sobre el vector completo de dimensión 9.

### *4.1 Análisis por país*

#### 4.1.1 Estados Unidos

A continuación se muestran los valores del AIC para modelos VAR para las 3 series estimados con distinto número de rezagos.

| Número de Rezagos | 1 | 2 | 3 |
|---|---|---|---|
| **AIC** | -1566.6 | -1599.5 | -1596.9 |

Usaremos un VAR de orden 2, sugerido por el AIC, para el análisis de cointegración. A continuación se muestran los resultados de la prueba de cointegración de las tasas de corto, mediano y largo plazo de Estados Unidos.

| H0 | r | Traza | Traza-95% |
|---|---|---|---|
| r = 0 | 0 | 25.7178 | 29.8 |
| r <= 1 | 1 | 8.6487 | 15.41 |
| r <= 2 | 2 | 0.8098 | 3.84 |

Se acepta la primer hipótesis nula que indica que el orden de cointegración es cero. Este primer resultado ya contrasta con los de Gonzalo y Granger para el periodo de 1969-1988 en los que el espacio de cointegración es de dos. Dado que no hay relaciones de cointegración entre las tasas de Estados Unidos no existe una descomposición P-T de su VAR.

## 4.1.2 Canadá

A continuación se muestran los valores del AIC para modelos VAR para las 3 series estimados con distinto número de rezagos.

| Número de Rezagos | 1 | 2 | 3 |
|---|---|---|---|
| **AIC** | -1563.3 | -1566.5 | -1578.5 |

Usaremos un VAR de orden 3, sugerido por el AIC, para el análisis de cointegración. A continuación se muestran los resultados de la prueba de cointegración de las tasas de corto, mediano y largo plazo de Canadá.

| H0 | r | Traza | Traza-95% |
|---|---|---|---|
| r = 0 | 0 | 34.2 | 29.8 |
| r <= 1 | 1 | 4.4 | 15.41 |
| r <= 2 | 2 | 0.6 | 3.84 |

En este caso se rechaza la hipótesis nula de que el espacio de cointegración es de orden cero y se acepta que es menor o igual a uno por lo que se concluye que la dimensión del espacio de cointegración es uno. Para el periodo estudiado por Gonzalo y Granger la dimensión del espacio de cointegración es dos. Parece que tanto en el caso de Estados Unidos como de Canadá el grado de complejidad del sistema, medido por el número de factores de largo plazo, ha incrementado. A continuación se presenta la descomposición P-T propuesta por Gonzalo y Granger.

|  | $\alpha$ | $\beta$ | $A_2$ |
|---|---|---|---|
|  | $z_1$ | $z_1$ | $z_1$ |
| $C_{cp}$ | 0.559 | 0.206 | 0.465 |
| $C_{Mp}$ | -0.772 | -0.455 | -0.748 |
| $C_{Lp}$ | -0.303 | 0.286 | -1.527 |

|  | $\alpha_\perp$ | | $\beta_\perp$ | | $A_1$ | |
|---|---|---|---|---|---|---|
|  | $f_1$ | $f_2$ | $f_1$ | $f_2$ | $f_1$ | $f_2$ |
| $C_{cp}$ | 0.059 | 0.308 | 0.791 | -0.497 | 0.465 | 2.176 |
| $C_{Mp}$ | 0.362 | 0.230 | 0.539 | 0.289 | -0.748 | 1.504 |
| $C_{Lp}$ | -0.814 | -0.019 | 0.289 | 0.818 | -1.527 | 0.827 |

A continuación se realizan pruebas de hipótesis para corroborar de qué variables del sistema dependen los dos factores permanentes encontrados. Como el espacio de tendencias comunes es de dimensión dos y hay tres variables en el sistema, los factores dependen de dos o tres de las tasas. Cada renglón de la tabla corresponde a la hipótesis nula de que los factores no dependen de la tasa de interés indicada. Por ejemplo, el primer renglón corresponde a la siguiente hipótesis nula:

$$\alpha_\perp = G\theta \quad \text{con} \ G = \begin{pmatrix} 0 & 0 \\ 1 & 0 \\ 0 & 1 \end{pmatrix}$$

| $G \in \mathbb{R}^{pxm}$ | | | Prueba $H_0: \alpha_\perp = G\theta$ | | | |
|---|---|---|---|---|---|---|
| Número de tasas que no influyen en factores | Número de tasas que componen 3 factores ($m$) | Tasas excluidas | Grados de libertad | Estadístico de prueba | valor crítico | valor-p |
| 1 | 2 | $C_{Lp}$ | 2 | 2.20E+01 | 5.991 | 0.000 |
| 1 | 2 | $C_{Mp}$ | 2 | 9.98E+00 | 5.991 | 0.007 |
| 1 | 2 | $C_{cp}$ | 2 | 9.69E+00 | 5.991 | 0.008 |

Se rechazan las hipótesis que los factores de largo plazo del sistema de Canadá no dependen de las tasas de corto, mediano y largo plazo. A continuación se muestran los dos factores permanentes, el componente transitorio y la descomposición de cada tasa en su componente transitorio y permanente.

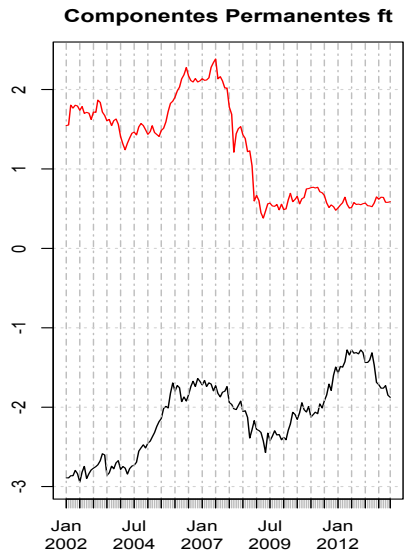
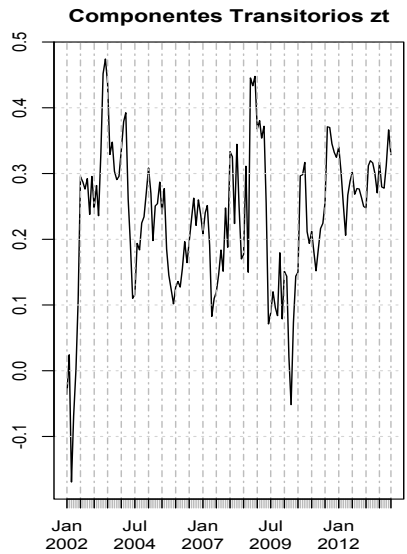
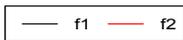
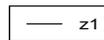

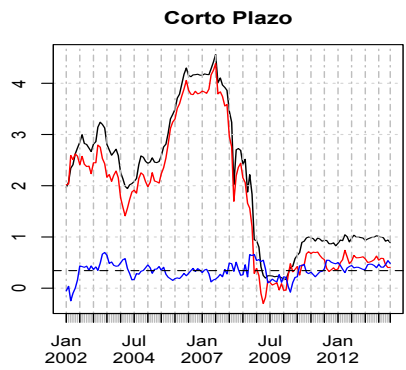
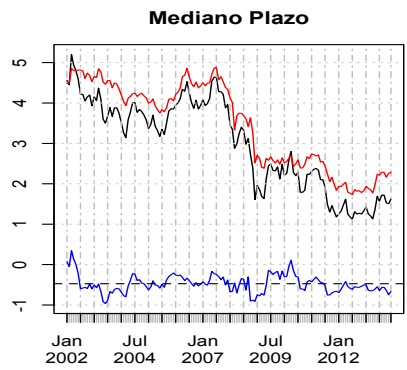
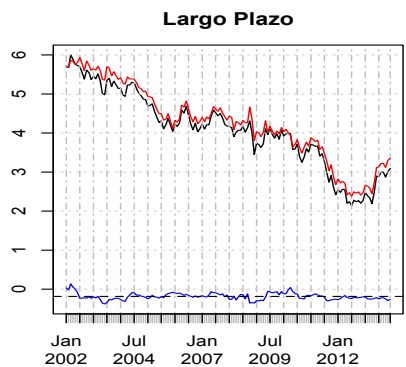
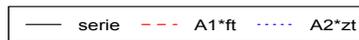

Cabe resaltar que lo que se grafica en la última figura son los componentes transitorios ($A_2 z_t$) y permanentes ($A_1 f_t$) de las tres tasas que se obtienen a partir de combinaciones lineales de los factores transitorios ($z_t$) y permanentes ($f_t$) que definen el espacio de cointegración y el espacio de tendencias comunes respectivamente:

$$X_t = A_1 f_t + A_2 \beta' X_t = A_1 f_t + A_2 z_t$$

### 4.1.3 México

A continuación se muestran los valores del AIC para modelos VAR para las 3 series estimados con distinto número de rezagos.

| Número de Rezagos | 1 | 2 | 3 |
|---|---|---|---|
| **AIC** | -840.0 | -845.4 | -848.0 |

Usaremos un VAR de orden 3, sugerido por el AIC, para el análisis de cointegración. A continuación se muestran los resultados de la prueba de cointegración de las tasas de corto, mediano y largo plazo de México.

| H0 | r | Traza | Traza-95% |
|---|---|---|---|
| r = 0 | 0 | 29.0 | 29.8 |
| r <= 1 | 1 | 8.6 | 15.41 |
| r <= 2 | 2 | 1.5 | 3.84 |

Se acepta la primer hipótesis nula que indica que el orden de cointegración es cero. Dado que no hay relaciones de cointegración entre las tasas de México no existe una descomposición P-T de su VAR.

### 4.1.4 Cointegración de factores por país

Del análisis por país determinamos que los sistemas de tasas de Estados Unidos, Canadá y México dependen, en el largo plazo, de tres, dos y tres factores respectivamente. Realizamos un análisis de cointegración sobre estos ocho factores para corroborar si el espacio de tendencias comunes de los tres países se puede reducir.

A continuación se muestran los valores del AIC para modelos VAR para las 8 series estimados con distinto número de rezagos.

| Número de Rezagos | 1 | 2 | 3 |
|---|---|---|---|
| **AIC** | -3991.9 | -4046.9 | -4018.4 |

Usaremos un VAR de orden 2, como sugiere el AIC, para el análisis de cointegración.

| H0 | r | Traza | Traza-95% |
|---|---|---|---|
| r = 0 | 0 | 186.1 | 159.3 |
| r <= 1 | 1 | 131.1 | 125.4 |
| r <= 2 | 2 | 81.1 | 95.5 |
| r <= 3 | 3 | 50.0 | 69.6 |
| r <= 4 | 4 | 23.3 | 47.7 |
| r <= 5 | 5 | 11.3 | 29.8 |
| r <= 6 | 6 | 3.9 | 15.4 |
| r <= 7 | 7 | 0.5 | 3.8 |

Los espacios de cointegración y de tendencias comunes son de dimensión dos y seis respectivamente. Es decir que en el largo plazo, el sistema de nueve tasas de Estados Unidos, Canadá y México depende de 6 factores. A continuación se presenta la descomposición P-T del VAR de Estados Unidos, Canadá y México.

|  | $\alpha$ | | $\beta$ | | $A_2$ | |
|---|---|---|---|---|---|---|
|  | $z_1$ | $z_2$ | $z_1$ | $z_2$ | $z_1$ | $z_2$ |
| $f_{E1}$ | 0.318 | -0.220 | 0.098 | 0.175 | -1.547 | 0.449 |
| $f_{E2}$ | 0.547 | 0.103 | 0.078 | -0.165 | -2.817 | -1.231 |
| $f_{E3}$ | 0.551 | 0.167 | -0.260 | 0.018 | -2.858 | -1.504 |
| $f_{C1}$ | 0.042 | -0.073 | -0.414 | -0.426 | -0.191 | 0.240 |
| $f_{C2}$ | 0.081 | 0.018 | -0.071 | -0.121 | -0.418 | -0.194 |
| $f_{M1}$ | -0.429 | 0.018 | 0.080 | -0.006 | 2.175 | 0.552 |
| $f_{M2}$ | 0.312 | 0.029 | -0.207 | 0.147 | -1.596 | -0.576 |
| $f_{M3}$ | -0.082 | 0.952 | 0.109 | -0.238 | 0.115 | -3.841 |

|  | $\alpha_\perp$ | | | | | |
|---|---|---|---|---|---|---|
|  | $f_1$ | $f_2$ | $f_3$ | $f_4$ | $f_5$ | $f_6$ |
| $f_{E1}$ | -0.301 | -0.151 | -0.291 | -0.324 | 0.190 | 0.151 |
| $f_{E2}$ | 0.528 | 0.033 | 0.141 | 0.334 | -0.050 | 0.658 |
| $f_{E3}$ | -0.396 | 0.173 | 0.118 | -0.418 | 0.180 | -0.636 |
| $f_{C1}$ | 0.143 | 0.319 | -0.488 | 0.191 | 0.453 | -0.967 |
| $f_{C2}$ | -0.350 | 0.509 | -0.263 | 0.621 | -0.733 | 0.089 |
| $f_{M1}$ | -0.090 | 0.157 | 0.090 | -0.046 | 0.157 | 0.033 |
| $f_{M2}$ | 0.018 | -0.182 | 0.063 | 0.221 | -0.061 | -0.031 |
| $f_{M3}$ | -0.038 | -0.052 | -0.139 | -0.041 | 0.065 | 0.000 |

|  | $\beta_\perp$ | | | | | |
|---|---|---|---|---|---|---|
|  | $f_1$ | $f_2$ | $f_3$ | $f_4$ | $f_5$ | $f_6$ |
| $f_{E1}$ | 0.367 | 0.824 | 0.167 | -0.113 | 0.222 | -0.032 |
| $f_{E2}$ | 0.372 | -0.237 | -0.128 | -0.115 | 0.539 | -0.575 |
| $f_{E3}$ | 0.779 | -0.244 | -0.031 | 0.068 | -0.208 | 0.147 |
| $f_{C1}$ | -0.150 | 0.408 | -0.139 | 0.046 | -0.016 | -0.116 |
| $f_{C2}$ | -0.005 | -0.124 | 0.966 | 0.001 | 0.030 | -0.057 |
| $f_{M1}$ | 0.068 | 0.075 | 0.009 | 0.979 | 0.064 | -0.046 |
| $f_{M2}$ | -0.235 | -0.135 | 0.002 | 0.073 | 0.743 | 0.219 |
| $f_{M3}$ | 0.194 | -0.001 | -0.033 | -0.060 | 0.246 | 0.762 |

|  | $A_1$ | | | | | |
|---|---|---|---|---|---|---|
|  | $f_1$ | $f_2$ | $f_3$ | $f_4$ | $f_5$ | $f_6$ |
| $f_{E1}$ | -0.473 | 0.287 | -1.465 | -0.699 | 0.602 | 1.373 |
| $f_{E2}$ | 0.196 | 0.626 | 0.002 | -1.001 | 0.198 | 1.878 |
| $f_{E3}$ | -0.671 | 0.302 | 0.690 | -1.513 | -0.341 | 1.078 |
| $f_{C1}$ | 0.447 | 0.405 | -0.685 | 0.250 | 0.475 | -0.241 |
| $f_{C2}$ | -0.515 | 0.478 | -0.279 | 0.181 | -0.420 | 0.326 |
| $f_{M1}$ | -2.141 | 0.802 | 1.663 | 2.893 | 3.125 | -0.031 |
| $f_{M2}$ | -2.060 | -1.918 | 1.051 | 2.175 | 1.436 | 0.569 |
| $f_{M3}$ | -2.298 | -2.379 | 0.956 | 0.803 | 0.457 | 0.407 |

Como el espacio de tendencias comunes es de dimensión seis y hay ocho variables en el sistema, los factores dependen de seis, siete u ocho de las variables. Dicho de otra forma en cada prueba de hipótesis se puede escoger, una o dos tasas a excluir para ver si al no rechazar la hipótesis nula corroboramos que el sistema no depende de alguna o algunas tasas. Se probaron todas las 28 posibilidades (combinaciones de 2 en 8). Únicamente se presentan resultados para las tres pruebas con valor-p más grande. Es decir, aquellas más lejanas de ser rechazadas.

| $G \in \mathbb{R}^{pxm}$ | | | Prueba $H_0: \alpha_\perp = G\theta$ | | | |
|---|---|---|---|---|---|---|
| Número de tasas que no influyen en factores | Número de tasas que componen 6 factores ($m$) | Tasas excluidas | Grados de libertad | Estadístico de prueba | valor crítico | valor-p |
| 1 | 7 | $f_{M2}$ | 6 | 2.88E+01 | 12.59159 | 0.000 |
| 1 | 7 | $f_{E1}$ | 6 | 2.29E+01 | 12.59159 | 0.001 |
| 1 | 7 | $f_{M3}$ | 6 | 2.05E+01 | 12.59159 | 0.002 |

A partir de las pruebas de hipótesis se concluye que el sistema depende de los ocho factores encontrados por medio del análisis por país. Como los dos factores de Canadá dependen de las tres tasas de interés del sistema de Canadá se concluye que, aún cuando existe cointegración de orden tres, las nueves tasas de interés impulsan al sistema. A continuación se muestran los dos factores permanentes, el componente transitorio y la descomposición de cada tasa en su componente transitorio y permanente.

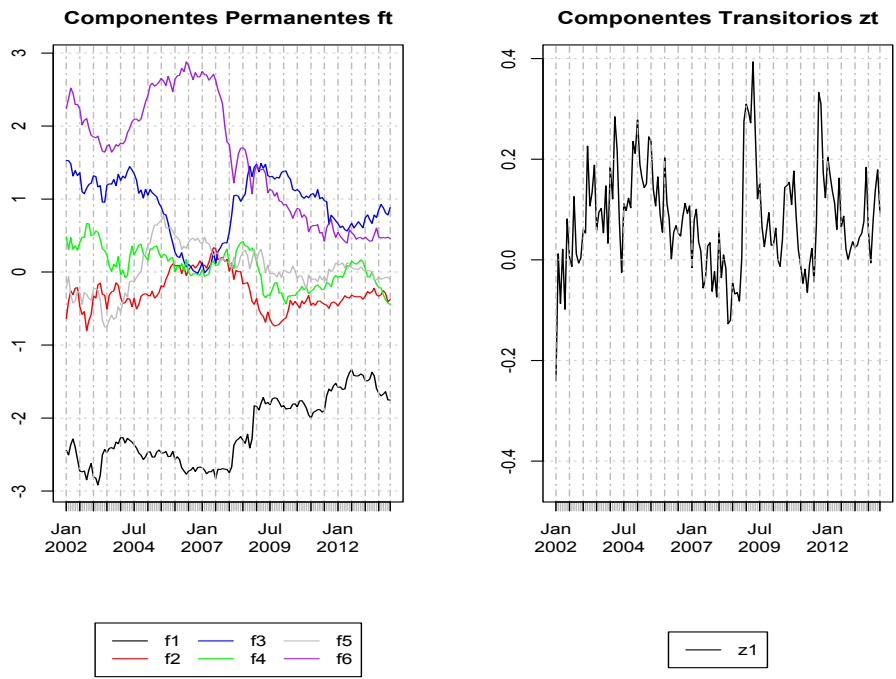

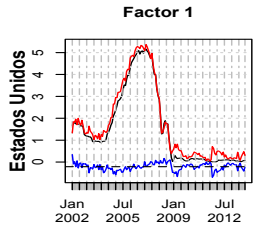
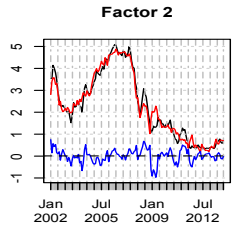
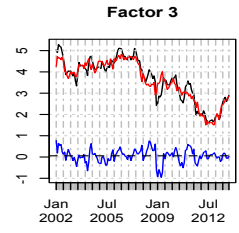
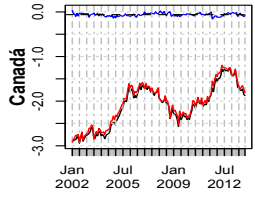
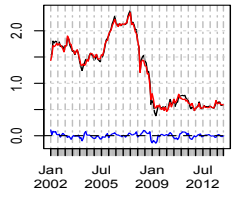
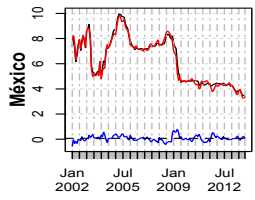
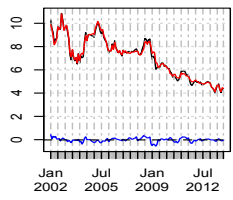
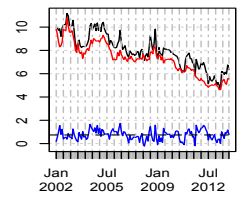

## 4.2 Análisis conjunto

Ahora realizamos el análisis de cointegración sobre el sistema de tasas de Estados Unidos, Canadá y México de forma conjunta. A continuación se muestran los valores del AIC para modelos VAR para las 9 series estimados con distinto número de rezagos.

| Número de Rezagos | 1 | 2 | 3 |
|---|---|---|---|
| **AIC** | -4257.5 | -4327.5 | -4285.9 |

Usaremos un VAR de orden 2, sugerido por el AIC, para el análisis de cointegración.

| H0 | r | Traza | Traza-95% |
|---|---|---|---|
| r = 0 | 0 | 261.2481 | 197.22 |
| r <= 1 | 1 | 183.9663 | 159.32 |
| r <= 2 | 2 | 129.2111 | 125.42 |
| r <= 3 | 3 | 83.5109 | 95.51 |
| r <= 4 | 4 | 52.4413 | 69.61 |
| r <= 5 | 5 | 25.5274 | 47.71 |
| r <= 6 | 6 | 10.9728 | 29.8 |
| r <= 7 | 7 | 3.6175 | 15.41 |
| r <= 8 | 8 | 0.2811 | 3.84 |

Los espacios de cointegración y de tendencias comunes del sistema conjunto tienen dimensión tres y seis respectivamente. En términos de la dimensión de los espacios de cointegración y de tendencias comunes llegamos al mismo resultado realizando el método por país o de forma conjunta. A continuación se presenta la descomposición P-T del VAR del sistema en componentes permanentes y transitorios.

|  | $\alpha$ | | |
|---|---|---|---|
|  | $z_1$ | $z_2$ | $z_3$ |
| $E_{CP}$ | -0.001 | 0.380 | -0.296 |
| $E_{MP}$ | -0.064 | 0.558 | -0.022 |
| $E_{LP}$ | -0.285 | 0.401 | 0.105 |
| $C_{cp}$ | -0.666 | -0.375 | 0.058 |
| $C_{Mp}$ | -0.002 | 0.419 | 0.137 |
| $C_{Lp}$ | -0.045 | 0.085 | 0.172 |
| $M_{CP}$ | 0.644 | -0.076 | -0.018 |
| $M_{MP}$ | -0.210 | 0.230 | 0.017 |
| $M_{LP}$ | 0.097 | 0.029 | 0.921 |

|  | $\beta$ | | |
|---|---|---|---|
|  | $z_1$ | $z_2$ | $z_3$ |
| $E_{CP}$ | 0.019 | 0.042 | 0.254 |
| $E_{MP}$ | -0.114 | 0.110 | -0.170 |
| $E_{LP}$ | -0.049 | -0.274 | -0.196 |
| $C_{cp}$ | -0.177 | -0.086 | -0.229 |
| $C_{Mp}$ | 0.630 | 0.082 | 0.127 |
| $C_{Lp}$ | -0.373 | 0.098 | 0.336 |
| $M_{CP}$ | 0.041 | 0.090 | 0.065 |
| $M_{MP}$ | -0.071 | -0.272 | -0.016 |
| $M_{LP}$ | 0.054 | 0.206 | -0.153 |

|  | $A_2$ | | |
|---|---|---|---|
|  | $z_1$ | $z_2$ | $z_3$ |
| $E_{CP}$ | 1.467 | -1.665 | 0.578 |
| $E_{MP}$ | 2.818 | -1.784 | -0.720 |
| $E_{LP}$ | 2.341 | -1.619 | -1.440 |
| $C_{cp}$ | -1.435 | -0.417 | -1.155 |
| $C_{Mp}$ | 2.324 | -0.927 | -0.979 |
| $C_{Lp}$ | 0.718 | -0.011 | -0.796 |
| $M_{CP}$ | -0.762 | 1.725 | 1.502 |
| $M_{MP}$ | 1.302 | -1.124 | -0.772 |
| $M_{LP}$ | 1.500 | 1.959 | -3.044 |

|  | $\alpha_\perp$ | | | | | |
|---|---|---|---|---|---|---|
|  | $f_1$ | $f_2$ | $f_3$ | $f_4$ | $f_5$ | $f_6$ |
| $E_{CP}$ | 0.388 | -0.138 | -0.240 | 0.285 | 0.054 | 0.046 |
| $E_{MP}$ | -0.590 | 0.017 | 0.124 | -0.347 | -0.025 | 0.720 |
| $E_{LP}$ | 0.325 | 0.084 | 0.332 | 0.476 | 0.326 | -0.534 |
| $C_{cp}$ | 0.029 | 0.155 | -0.059 | -0.082 | 0.044 | 0.147 |
| $C_{Mp}$ | 0.172 | 0.350 | -0.353 | -0.220 | -0.142 | -0.486 |
| $C_{Lp}$ | 0.047 | -0.350 | 0.363 | 0.076 | -0.444 | 0.763 |
| $M_{CP}$ | 0.103 | 0.127 | 0.137 | 0.015 | 0.121 | 0.022 |
| $M_{MP}$ | -0.032 | -0.168 | -0.010 | -0.224 | -0.071 | -0.045 |
| $M_{LP}$ | 0.040 | -0.044 | -0.121 | 0.057 | 0.085 | 0.014 |

|  | $\beta_\perp$ | | | | | |
| --- | --- | --- | --- | --- | --- | --- |
|  | $f_1$ | $f_2$ | $f_3$ | $f_4$ | $f_5$ | $f_6$ |
| $E_{CP}$ | 0.311 | -0.774 | 0.254 | -0.100 | 0.181 | -0.075 |
| $E_{MP}$ | -0.120 | 0.032 | 0.193 | -0.106 | 0.555 | -0.631 |
| $E_{LP}$ | -0.274 | -0.097 | 0.709 | 0.137 | -0.211 | -0.053 |
| $C_{cp}$ | 0.873 | 0.141 | 0.106 | 0.051 | -0.083 | 0.002 |
| $C_{Mp}$ | 0.211 | 0.360 | 0.310 | -0.065 | 0.132 | -0.086 |
| $C_{Lp}$ | 0.018 | 0.460 | 0.409 | -0.034 | 0.031 | 0.086 |
| $M_{CP}$ | 0.023 | -0.040 | -0.006 | 0.974 | 0.085 | -0.071 |
| $M_{MP}$ | 0.017 | 0.112 | -0.157 | 0.057 | 0.732 | 0.296 |
| $M_{LP}$ | -0.079 | -0.126 | 0.310 | -0.025 | 0.215 | 0.697 |

|  | $A_1$ | | | | | |
| --- | --- | --- | --- | --- | --- | --- |
|  | $f_1$ | $f_2$ | $f_3$ | $f_4$ | $f_5$ | $f_6$ |
| $E_{CP}$ | 0.397 | -0.312 | -1.115 | 1.005 | 1.264 | 1.675 |
| $E_{MP}$ | -0.451 | -0.264 | 0.238 | 0.954 | 0.674 | 2.015 |
| $E_{LP}$ | 0.409 | -0.327 | 0.946 | 1.323 | -0.251 | 1.084 |
| $C_{cp}$ | 1.344 | 1.852 | -0.802 | -0.350 | -0.830 | 0.561 |
| $C_{Mp}$ | 0.760 | 0.257 | 0.343 | 0.289 | -1.082 | 0.936 |
| $C_{Lp}$ | 0.921 | -0.247 | 0.857 | 0.190 | -1.436 | 0.580 |
| $M_{CP}$ | 2.073 | 0.124 | 1.794 | -3.326 | 2.938 | -0.139 |
| $M_{MP}$ | 1.780 | -2.353 | 0.768 | -2.423 | 1.103 | 0.449 |
| $M_{LP}$ | 1.973 | -2.603 | 0.715 | -1.057 | -0.011 | 0.264 |

Realizando el análisis por país llegamos a la conclusión que las nueve tasas de interés impulsan al sistema. Veamos si llegamos a conclusiones similares con el análisis conjunto. En este caso la dimensión del espacio de tendencias comunes es seis y el número total de variables es nueve. Esto significa que los factores se pueden componer de seis, siete, ocho o nueve variables. Dicho de otra forma en cada prueba de hipótesis se puede escoger, una, dos o tres tasas a excluir para ver si al no rechazar la hipótesis nula corroboramos que el sistema no depende de alguna o algunas tasas. Se probaron todas las 84 posibilidades (combinaciones de 3 en 9). Únicamente se presentan resultados para las tres pruebas con valor p más grande. Es decir, aquellas más lejanas de ser rechazadas.

| $G \in \mathbb{R}^{pxm}$ | | | Prueba $H_0: \alpha_\perp = G\theta$ | | | |
|---|---|---|---|---|---|---|
| Número de tasas que no influyen en factores | Número de tasas que componen 6 factores ($m$) | Tasas excluidas | Grados de libertad | Estadístico de prueba | valor crítico | valor-p |
| 1 | 8 | $M_{LP}$ | 6 | 2.10E+01 | 12.592 | 0.002 |
| 2 | 7 | $C_{cp}, M_{LP}$ | 12 | 2.99E+01 | 21.026 | 0.003 |
| 1 | 8 | $E_{CP}$ | 6 | 1.40E+01 | 12.592 | 0.030 |
| 1 | 8 | $C_{cp}$ | 6 | 8.26E+00 | 12.592 | 0.220 |

De acuerdo a estas pruebas de hipótesis la única tasa que no impulsa al sistema en el largo plazo es la de corto plazo de Canadá. A continuación se muestran los dos factores permanentes, el componente transitorio y la descomposición de cada tasa en su componente transitorio y permanente.

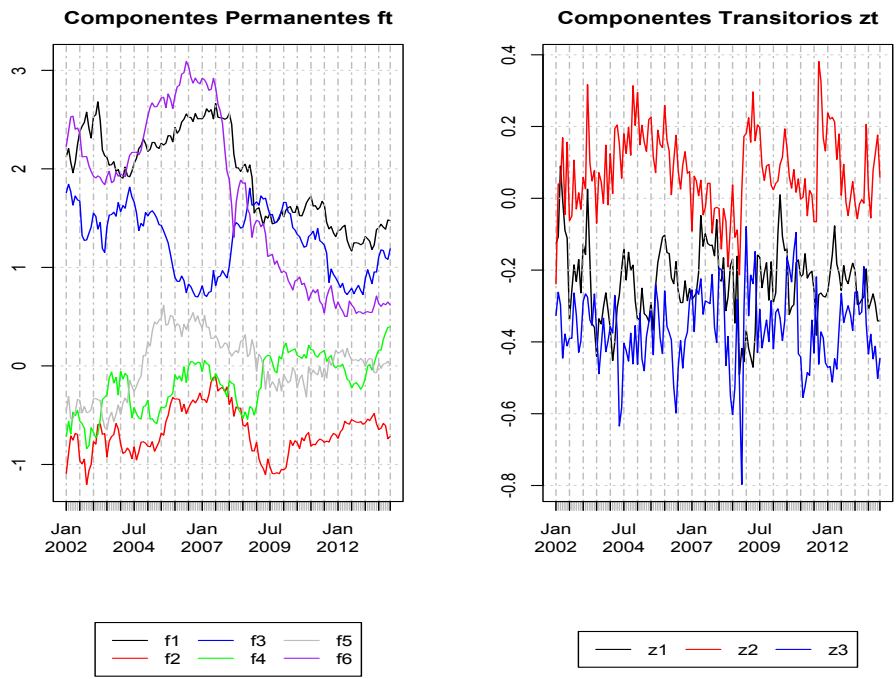

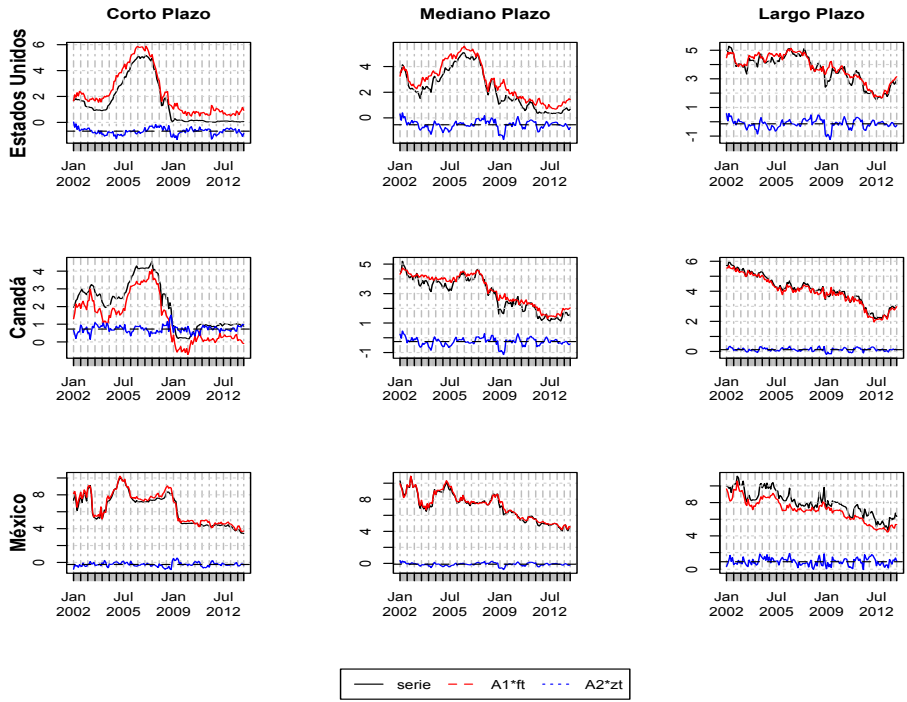

## *4.3 Conclusiones*

Para el periodo de 1969-1988 Gonzalo y Granger observaron que todo el sistema de tasas de Estados Unidos y Canadá dependía de un solo factor. Además comprobaron que este factor se compone únicamente por las tasas de Estados Unidos. Para el periodo de 2002-2013 se observó un orden de cointegración mucho menor. Para el sistema de Estados Unidos no se observó relación de cointegración alguna y para Canadá el orden de cointegración es de uno, cuando en el periodo de 1969-1988 se observan órdenes de cointegración de dos en ambos sistemas. Además se encontró que los seis factores permanentes que impulsan el sistema se componen de las nueve tasas excepto, posiblemente, la de corto plazo de Canadá.

Una posible explicación es que en realidad el espacio de tendencias comunes no ha incrementado pero los factores permanentes son más complejos y ya no son una función lineal de las tasas de interés. En este caso , la descomposición P-T de Gonzalo y Granger no sería adecuada para identificar los componentes permanente y transitorio. El hecho de que el periodo analizado incluye la crisis financiera de 2007-2008 le da credibilidad a esta hipótesis puesto que las relaciones de los mercados en épocas de crisis se vuelven más complejas.

Por otro lado, puede ser que si existan seis factores permanentes. Esto se podría explicar a partir sistemas financieros en Canadá y México que han crecido en tamaño y sofisticación y que ya no dependen de la misma manera de lo que sucede en Estados Unidos.

# 5   Bibliografía

## 6  Anexo (Código en RATS)

A continuación se anexa el código en RATS con el que se realizó el análisis de cointegración.

```
*************************************************************************
*
* Análisis de Cointegración de las tasas de interés de Estados Unidos,
* México y Canadá. Análogo al estudio realizdo por Gonzalo et al. (1995)
* peropara el periodo de enero de 2002 a diciembre de 2013.
*************************************************************************
*
* Se cargan datos y se declara el periodo (mensual) de estudio.
open data usaCanMex.rat
calendar(m) 2002:1
data(format=rats) 2002:1 2013:12

*************************************************************************
*
* Análisis exploratorio
*************************************************************************
*
* Se grafican las tasas de inters de corto, mediano y largo plazo,
* respectivamente,de Estados Unidos, Canadá y México, respectivamente.
graph(footer="Gráfica 1 Tasas de interés Estados Unidos",key=upleft) 3
# USAX3m
# USAX3y
# USAX10y
graph(footer="Gráfica 2 Tasas de interés Canadá",key=upleft) 3
# CANX3m
# CANX3y
# CANX10y
graph(footer="Gráfica 3 Tasas de interés México",key=upleft) 3
# MEXX3m
# MEXX3y
# MEXX10y
*************************************************************************
*
```

* Pruebas Dickey-Fuller Aumentadas
**************************************************************************
*
* Se carga paquete con rutina para realizar prueba Dickey-Fuller Aumentada
source dfunit.src
* Para cada una de las 9 series se realiza la prueba Dickey-Fuller Aumentada
* para corroborar la existencia de raices unitarias en el vector.
report(action=define)
report(atcol=2) "ADF(0)" "ADF(1)" "ADF(2)" "ADF(3)" "ADF(4)"
dofor s = USAX3m USAX3y USAX10y CANX3m CANX3y CANX10y MEXX3m MEXX3y MEXX10y
   report(row=new,atcol=1) %L(s)
   do lags=0,4
      @dfunit(lags=lags,noprint) s 2002:5-lags
      report(row=current,atcol=lags+2) %cdstat
   end do lags
end dofor s
report(action=format,picture="*.##")
report(action=show)

**************************************************************************
*
* Orden del VAR
**************************************************************************
*

***************************
*Modelo VAR ESTADOS UNIDOS
***************************

* Se checaran modelos con 1,2,3 rezagos
COMPUTE FIXLAG = 3

*Se definen modelos VAR con distinto numero de rezagos
COMPUTE p=3              ;* Número de variables
DO I = 1,3,1
```



```
SMPL (2002:01+FIXLAG) 2013:12

LINREG(NOPRINT,DEFINE=EQ1) USAX3M /
# constant USAX3m{1 TO I} USAX3y{1 TO I} USAX10y{1 TO I}

LINREG(NOPRINT,DEFINE=EQ2) USAX3y /
# constant USAX3m{1 TO I} USAX3y{1 TO I} USAX10y{1 TO I}

LINREG(NOPRINT,DEFINE=EQ3) USAX10y /
# constant USAX3m{1 TO I} USAX3y{1 TO I} USAX10y{1 TO I}

* Se estima la matriz de varianzas y covarianzas para el vector de errores
* correspondiente del modelo con I rezagos
SUR(NOPRINT,OUTSIGMA=UCOV,ITER=0) 3 2002:01+FIXLAG 2013:12
#EQ1
#EQ2
#EQ3

* Logaritmo natural del determinante de sigma
COMPUTE LNDETCOV=LOG(%DET(UCOV))
WRITE LNDETCOV

* Se calcula el AIC y SBC para el modelo con k=I rezagos
COMPUTE k = I
EVAL AIC = (%NOBS)*LNDETCOV + 2.0*((p**2)*k+p)
EVAL SBC = (%NOBS)*LNDETCOV + ((p**2)*k+p)*LOG(%NOBS)

*EVAL AIC = (%NOBS-P)*LNDETCOV + 2.0*((p**2)*k+p)
*EVAL SBC = (%NOBS-P)*LNDETCOV + ((p**2)*k+p)*LOG(%NOBS-P)

* Se imprime el AIC y SBC para el modelo con k=I rezagos
DISPLAY 'LAG=' k
DISPLAY 'AIC-VALUE=' AIC
DISPLAY 'SBC-VALUE=' SBC
```



```
DISPLAY 'NOBS = ' %NOBS
END DO

***************************
*Modelo VAR CANADÁ
***************************

* Se checaran modelos con 1,2,3 rezagos
COMPUTE FIXLAG = 3

*Se definen modelos VAR con distinto numero de rezagos
COMPUTE p=3              ;* Número de variables
DO I = 1,3,1
SMPL (2002:01+FIXLAG) 2013:12

LINREG(NOPRINT,DEFINE=EQ1) CANX3m /
# constant CANX3m{1 TO I} CANX3y{1 TO I} CANX10y{1 TO I}

LINREG(NOPRINT,DEFINE=EQ2) CANX3y /
# constant CANX3m{1 TO I} CANX3y{1 TO I} CANX10y{1 TO I}

LINREG(NOPRINT,DEFINE=EQ3) CANX10y /
# constant CANX3m{1 TO I} CANX3y{1 TO I} CANX10y{1 TO I}

* Se estima la matriz de varianzas y covarianzas para el vector de errores
* correspondiente del modelo con I rezagos
SUR(NOPRINT,OUTSIGMA=UCOV,ITER=0) 3 2002:01+FIXLAG 2013:12
#EQ1
#EQ2
#EQ3

* Logaritmo natural del determinante de sigma
COMPUTE LNDETCOV=LOG(%DET(UCOV))
WRITE LNDETCOV
```



* Se calcula el AIC y SBC para el modelo con k=I rezagos
COMPUTE k = I
EVAL AIC = (%NOBS)*LNDETCOV + 2.0*((p**2)*k+p)
EVAL SBC = (%NOBS)*LNDETCOV + ((p**2)*k+p)*LOG(%NOBS)

*EVAL AIC = (%NOBS-P)*LNDETCOV + 2.0*((p**2)*k+p)
*EVAL SBC = (%NOBS-P)*LNDETCOV + ((p**2)*k+p)*LOG(%NOBS-P)

* Se imprime el AIC y SBC para el modelo con k=I rezagos
DISPLAY 'LAG=' k
DISPLAY 'AIC-VALUE=' AIC
DISPLAY 'SBC-VALUE=' SBC
DISPLAY 'NOBS = ' %NOBS
END DO

***************************
*Modelo VAR MÉXICO
***************************

* Se checaran modelos con 1,2,3 rezagos
COMPUTE FIXLAG = 3

*Se definen modelos VAR con distinto numero de rezagos
COMPUTE p=3             ;* Número de variables
DO I = 1,3,1
SMPL (2002:01+FIXLAG) 2013:12

LINREG(NOPRINT,DEFINE=EQ1) MEXX3m /
# constant MEXX3m{1 TO I} MEXX3y{1 TO I} MEXX10y{1 TO I}

LINREG(NOPRINT,DEFINE=EQ2) MEXX3y /
# constant MEXX3m{1 TO I} MEXX3y{1 TO I} MEXX10y{1 TO I}

LINREG(NOPRINT,DEFINE=EQ3) MEXX10y /
# constant MEXX3m{1 TO I} MEXX3y{1 TO I} MEXX10y{1 TO I}



```
* Se estima la matriz de varianzas y covarianzas para el vector de errores
* correspondiente del modelo con I rezagos
SUR(NOPRINT,OUTSIGMA=UCOV,ITER=0) 3 2002:01+FIXLAG 2013:12
#EQ1
#EQ2
#EQ3

* Logaritmo natural del determinante de sigma
COMPUTE LNDETCOV=LOG(%DET(UCOV))
WRITE LNDETCOV

* Se calcula el AIC y SBC para el modelo con k=I rezagos
COMPUTE k = I
EVAL AIC = (%NOBS)*LNDETCOV + 2.0*((p**2)*k+p)
EVAL SBC = (%NOBS)*LNDETCOV + ((p**2)*k+p)*LOG(%NOBS)

*EVAL AIC = (%NOBS-P)*LNDETCOV + 2.0*((p**2)*k+p)
*EVAL SBC = (%NOBS-P)*LNDETCOV + ((p**2)*k+p)*LOG(%NOBS-P)

* Se imprime el AIC y SBC para el modelo con k=I rezagos
DISPLAY 'LAG=' k
DISPLAY 'AIC-VALUE=' AIC
DISPLAY 'SBC-VALUE=' SBC
DISPLAY 'NOBS = ' %NOBS
END DO

***************************
*Modelo VAR COMPLETO
***************************

* Se checaran modelos con 1,2,3 rezagos
COMPUTE FIXLAG = 3

*Se definen modelos VAR con distinto numero de rezagos
COMPUTE p=3              ;* Número de variables
```



```
DO I = 1,3,1
SMPL (2002:01+FIXLAG) 2013:12

LINREG(NOPRINT,DEFINE=EQ1) USAX3M /
# constant USAX3m{1 TO I} USAX3y{1 TO I} USAX10y{1 TO I} CANX3m{1 TO I}
CANX3y{1 TO I} CANX10y{1 TO I} MEXX3m{1 TO I} MEXX3y{1 TO I} MEXX10y{1 TO I}

LINREG(NOPRINT,DEFINE=EQ2) USAX3y /
# constant USAX3m{1 TO I} USAX3y{1 TO I} USAX10y{1 TO I} CANX3m{1 TO I}
CANX3y{1 TO I} CANX10y{1 TO I} MEXX3m{1 TO I} MEXX3y{1 TO I} MEXX10y{1 TO I}

LINREG(NOPRINT,DEFINE=EQ3) USAX10y /
# constant USAX3m{1 TO I} USAX3y{1 TO I} USAX10y{1 TO I} CANX3m{1 TO I}
CANX3y{1 TO I} CANX10y{1 TO I} MEXX3m{1 TO I} MEXX3y{1 TO I} MEXX10y{1 TO I}

LINREG(NOPRINT,DEFINE=EQ4) CANX3m /
# constant USAX3m{1 TO I} USAX3y{1 TO I} USAX10y{1 TO I} CANX3m{1 TO I}
CANX3y{1 TO I} CANX10y{1 TO I} MEXX3m{1 TO I} MEXX3y{1 TO I} MEXX10y{1 TO I}

LINREG(NOPRINT,DEFINE=EQ5) CANX3y /
# constant USAX3m{1 TO I} USAX3y{1 TO I} USAX10y{1 TO I} CANX3m{1 TO I}
CANX3y{1 TO I} CANX10y{1 TO I} MEXX3m{1 TO I} MEXX3y{1 TO I} MEXX10y{1 TO I}

LINREG(NOPRINT,DEFINE=EQ6) CANX10y /
# constant USAX3m{1 TO I} USAX3y{1 TO I} USAX10y{1 TO I} CANX3m{1 TO I}
CANX3y{1 TO I} CANX10y{1 TO I} MEXX3m{1 TO I} MEXX3y{1 TO I} MEXX10y{1 TO I}

LINREG(NOPRINT,DEFINE=EQ7) MEXX3m /
# constant USAX3m{1 TO I} USAX3y{1 TO I} USAX10y{1 TO I} CANX3m{1 TO I}
CANX3y{1 TO I} CANX10y{1 TO I} MEXX3m{1 TO I} MEXX3y{1 TO I} MEXX10y{1 TO I}

LINREG(NOPRINT,DEFINE=EQ8) MEXX3y /
# constant USAX3m{1 TO I} USAX3y{1 TO I} USAX10y{1 TO I} CANX3m{1 TO I}
CANX3y{1 TO I} CANX10y{1 TO I} MEXX3m{1 TO I} MEXX3y{1 TO I} MEXX10y{1 TO I}

LINREG(NOPRINT,DEFINE=EQ9) MEXX10y /
```



```
# constant USAX3m{1 TO I} USAX3y{1 TO I} USAX10y{1 TO I} CANX3m{1 TO I}
CANX3y{1 TO I} CANX10y{1 TO I} MEXX3m{1 TO I} MEXX3y{1 TO I} MEXX10y{1 TO I}

* Se estima la matriz de varianzas y covarianzas para el vector de errores
* correspondiente del modelo con I rezagos
SUR(NOPRINT,OUTSIGMA=UCOV,ITER=0) 9 2002:01+FIXLAG 2013:12
#EQ1
#EQ2
#EQ3
#EQ4
#EQ5
#EQ6
#EQ7
#EQ8
#EQ9

* Logaritmo natural del determinante de sigma
COMPUTE LNDETCOV=LOG(%DET(UCOV))
WRITE LNDETCOV

* Se calcula el AIC y SBC para el modelo con k=I rezagos
COMPUTE k = I
EVAL AIC = (%NOBS)*LNDETCOV + 2.0*((p**2)*k+p)
EVAL SBC = (%NOBS)*LNDETCOV + ((p**2)*k+p)*LOG(%NOBS)

*EVAL AIC = (%NOBS-P)*LNDETCOV + 2.0*((p**2)*k+p)
*EVAL SBC = (%NOBS-P)*LNDETCOV + ((p**2)*k+p)*LOG(%NOBS-P)

* Se imprime el AIC y SBC para el modelo con k=I rezagos
DISPLAY 'LAG=' k
DISPLAY 'AIC-VALUE=' AIC
DISPLAY 'SBC-VALUE=' SBC
DISPLAY 'NOBS = ' %NOBS
END DO
```



*************************************************************************
*******
* Se realiza el procedimiento de Johansen para determinar el rango de la matriz pi
* en el modelo de corrección de errores y para determina alfa y beta tal que
* pi = alfa*t(beta)
*************************************************************************
*******
SMPL (2002:01) 2013:12

*Se realiza procedimient de johansen para sistema "Estados Unidos"
@johmle(lags=2,vectors=eusa,dual=dusa,eigenval=vusa, loadings=lusa)
# USAX3m USAX3y USAX10y
compute s00usa = %%s00
compute s11usa = %%s11
compute s01usa = %%s01

*Se realiza procedimient de johansen para sistema "Canadá"
@johmle(lags=3,vectors=ecanada,dual=dcanada,eigenval=vcanada, loadings=lcanada)
# CANX3m CANX3y CANX10y
compute s00can = %%s00
compute s11can = %%s11
compute s01can = %%s01

*Se realiza procedimient de johansen para sistema "México"
@johmle(lags=3,vectors=emexico,dual=dmexico,eigenval=vmexico, loadings=lmexico)
# MEXX3m MEXX3y MEXX10y
compute s00mex = %%s00
compute s11mex = %%s11
compute s01mex = %%s01

*Se realiza procedimient de johansen para sistema completo
@johmle(lags=2,vectors=ejoint,dual=djoint,eigenval=vjoint, loadings=ljoint)
# USAX3m USAX3y USAX10y CANX3m CANX3y CANX10y MEXX3m MEXX3y MEXX10y
compute s00joint = %%s00
compute s11joint = %%s11
compute s01joint = %%s01



```
********************************************************************
*****
* Se imprimen los distintos elementos necesarios para estimar alfa,
* beta, alfa-ortogonal y beta-ortogonal
********************************************************************
*****
disp "Numero Observaciones"
disp %nobs

* Para el sistema "Estados Unidos"
disp "United States"
disp "Eigenvalues" #.############### vusa
disp "Eigenvectors (W)"
disp ##.############### eusa/sqrt(%nobs)
disp "Eigenvectors (Z)"
disp ##.############### dusa/sqrt(%nobs)
disp "s00, s11, s01"
disp ##.############### s00usa
disp ##.############### s11usa
disp ##.############### s01usa

* Para el sistema "Canadá"
disp "Canada"
disp "Eigenvalues" #.############### vcanada
disp "Eigenvectors (W)"
disp ##.############### ecanada/sqrt(%nobs)
disp "Eigenvectors (Z)"
disp ##.############### dcanada/sqrt(%nobs)
disp "s00, s11, s01"
disp ##.############### s00can
disp ##.############### s11can
disp ##.############### s01can

* Para el sistema "México"
disp "Mexico"
```



```
disp "Eigenvalues" #.############## vmexico
disp "Eigenvectors (W)"
disp ##.############## emexico/sqrt(%nobs)
disp "Eigenvectors (Z)"
disp ##.############## dmexico/sqrt(%nobs)
disp "s00, s11, s01"
disp ##.############## s00mex
disp ##.############## s11mex
disp ##.############## s01mex

* Para el sistema completo
disp "Joint Analysis"
disp "Eigenvalues" #.############## vjoint
disp "Eigenvectors (W)"
disp ##.############## ejoint/sqrt(%nobs)
disp "Eigenvectors (Z)"
disp ##.############## djoint/sqrt(%nobs)
disp "s00, s11, s01"
disp ##.############## s00joint
disp ##.############## s11joint
disp ##.############## s01joint

************************************************************************
* Estimacion de alfa, alfa-ortogonal, beta, beta-ortogonal,
* A1 y A2 para el sistema completo
************************************************************************
*se guarda tamaño p del vector y dimensión r del espacio de cointegración
compute r=6,p=9

*******************
* ecuación auxiliar
*******************
equation xeqn *
# USAX3m USAX3y USAX10y CANX3m CANX3y CANX10y MEXX3m MEXX3y MEXX10y

*********************************************
```



```
* se calcula alfa, alfa-ortogonal, beta y beta-ortogonal
***********************************************************

compute alfaPerp=%xsubmat(djoint,1,p,r+1,p)
compute alfa=%perp(alfaPerp)
compute beta=%xsubmat(ejoint,1,p,1,r)
compute betaPerp=%perp(beta)

disp "alfa"
disp ##.############### alfa
disp "alfa perp"
disp ##.############### alfaPerp
disp "beta"
disp ##.############### beta
disp "beta_perp"
disp ##.############### betaPerp

***************************************************
* Se calculan matrices A1 y A2
***************************************************
compute a1=betaPerp*inv(tr(alfaPerp)*betaPerp)
compute a2=alfa*inv(tr(beta)*alfa)

disp "A1"
disp ##.############### a1
disp "A2"
disp ##.############### a2

*****************************************************
* GRAFICAS DE LA DESCOMPOSICION
*****************************************************
* Solo se muestran dos con fin de ilustración.
********************************************************************************
*****
* Descomposición de tasa de interés de corto plazo de Estados Unidos (variable 1)
```



```
************************************************************************
*****
compute pweight=%xrow(a1*tr(alfaPerp),1)
set pcomp = %dot(pweight,%eqnxvector(xeqn,t))
compute tweight=%xrow(a2*tr(beta),1)
set tcomp = %dot(tweight,%eqnxvector(xeqn,t))
graph(footer="Estados Unidos: Descomposición tasa corto plazo",key=upleft) 3
# USAX3m
# pcomp
# tcomp
************************************************************************
****
* Descomposición de tasa de interés de mediano plazo de México (variable 8)
************************************************************************
****
compute pweight=%xrow(a1*tr(alfaPerp),8)
set pcomp = %dot(pweight,%eqnxvector(xeqn,t))
compute tweight=%xrow(a2*tr(beta),8)
set tcomp = %dot(tweight,%eqnxvector(xeqn,t))
graph(footer="México: Descomposición de tasa mediano plazo",key=upleft) 3
# MEXX3y
# pcomp
# tcomp

*************************************************
* Pruebas de Hipotesis: alfa ortogonal = G*teta
*************************************************

************************************************************************
* Prueba que las tendencias de largo plazo solo dependen de tasas
* de estados unidos.
************************************************************************
compute m=3
dec rect g(p,m)
ewise g(i,j)=(i==j+3)
compute g00g=tr(g)*%%s00*g
```



```
compute g11g=tr(g)*%%s01*inv(%%s11)*tr(%%s01)*g
compute sf=inv(%decomp(g00g))
eigen sf*g11g*tr(sf) eigval eigvec
compute eigvec=tr(sf)*eigvec
disp "Constrained Weights" g*%xcol(eigvec,m)/sqrt(%nobs)
compute %cdstat=-%nobs*log((1-eigval(m))/(1-vjoint(p)))
cdf(title="H0:alfa-ort solo depende de tasas de USA") chisqr %cdstat 6-m
```